\documentclass[preprint,showpacs,preprintnumbers,amsmath,amssymb]{revtex4}

\usepackage{graphicx}
\usepackage{dcolumn}
\usepackage{bm}
\usepackage{multirow}

\usepackage{color}

\begin{document}

\preprint{APS/123-QED}

\title{Analytical considerations for linear and nonlinear optimization of the TME cells. Application to the CLIC pre-damping rings.}

\author{F. Antoniou}
 \email{Fanouria.Antoniou@cern.ch}
\affiliation{CERN}
\author{Y. Papaphilippou}%
 \email{Ioannis.Papaphilippou@cern.ch}
\affiliation{CERN}%

\date{\today}

\begin{abstract}

  The theoretical minimum emittance cells are the optimal configurations for achieving 
  the absolute minimum emittance, if specific optics constraints are satisfied at the 
  middle of the cell's dipole. 
  Linear lattice design options based on an analytical
  approach for the theoretical minimum emittance cells are presented in this paper.
  In particular the parametrization of the quadrupole strengths and
  optics functions with respect to the emittance and drift lengths is derived.
  A multi-parametric space can be then created with all the cell parameters, 
  from which one can chose any of them to be optimized. 
  An application of this approach are finally 
  presented for the linear and non-linear optimization of the CLIC Pre-damping rings.
  
\end{abstract}

\pacs{PAC09:WE6PFP107,EPAC08:MOPP062}
\maketitle

\section{\label{sec:level1}Introduction\protect\\ }

High brilliance or luminosity requirements, for electron storage or linear colliders' damping rings, 
necessitate ultra low emittance beams. Under the influence of synchrotron radiation, the theoretical 
minimum emittance (TME)~\cite{MinimumEmitRadiation}, is reached for specific optics conditions, 
including a unique high cell phase advance~\cite{PhysRevSTAB.4.021001}. The strong focusing needed for accomplishing the 
TME conditions results in cells with intrinsically high chromaticity. The chromatic sextupoles' 
strengths are enhanced by the low dispersion of the TME cell and reduce the Dynamic Aperture (DA). 
The ultimate target of a low emittance cell designer is to build a compact ring, attaining a sufficiently 
low emittance, with an adequately large DA, driven by geometrical aperture and injection requirements. 
The lattice design, however, is often based on numerical tools whose optimization algorithms depend 
heavily on the initial conditions. Reaching the optimal solution necessitates several iterations, 
without necessarily having a global  understanding of the interdependence between a series of optics 
parameters and knobs. Modern techniques, as the Multi-Objective Genetic Algorithms 
(MOGA)~\cite{ref:MOGA} or the Global Analysis of Stable Solutions (GLASS)~\cite{ref:GLASS} attempt to 
achieve a global optics optimization exploring numerically all possible solutions, within stability and 
performance requirements. In this paper, a different approach is followed, by obtaining an analytical 
solution for the quadrupole strengths and a complete parametrization of the TME cell, using thin lens 
approximation. In this way, all cell properties are globally determined and the optimization procedure 
following any design requirement can be performed in a systematic way. Although approximate, the 
obtained solutions are very close to the real thick-element optics and can be used as initial conditions 
for efficiently matching the lattice through numerical optics codes. 

The CLIC pre-damping rings offer an ideal test-bed for applying the procedure mentioned above: they have 
to accommodate a large  emittance beam, coming in particular from the positron source and
reduce its size to low enough values for injection into the main damping rings. The latter requirement 
imposes a low emittance cell linear optics design, whereas the former one necessitates a large off-momentum 
DA.

The paper is organized as follows: In Section~\ref{sec:AnalApproach}, the analytical expressions for the 
quadrupole strengths and other optics parameters of the TME cell are derived, including conditions for 
stability of the solutions and feasibility of the magnets. In Section~\ref{sec:NumApplication}, the 
complete parametrization of the TME cells is performed using numerical examples of the analytical 
thin-lens solutions, applicable to the CLIC Pre-damping rings (PDR) lattice design. A validation of the method 
through the comparison of the results with numerical simulations using MADX~\cite{bib:MADX} is presented in 
section~\ref{sec:MADXComp}. Finally, in section~\ref{sec:PDRDesign} an application of the analytical approach and the 
resonance free lattice concept~\cite{ref:ResFreeLat} is used for the linear and non-linear optimization of the CLIC PDR.
 
\section{\label{sec:AnalApproach} Analytical approach for the TME cells \protect\\ }

\subsection{The TME cell}

\begin{figure}[ht]
\begin{center}
\includegraphics*[width=100mm]{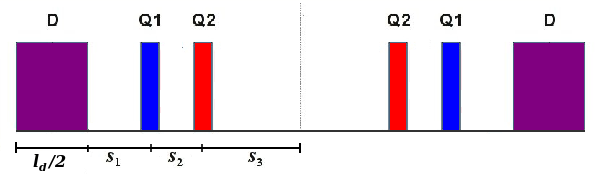}
\vspace{-10px}
\caption{\label{fig:minemg} Schematic layout of the TME cell.}
\end{center}
\end{figure}

A schematic layout of the TME cell is displayed in Fig.~\ref{fig:minemg}. It consists of one dipole D of 
length $l_d$ and at least two families of quadrupoles Q1, Q2, as pictured. The quadrupole focal lengths are 
denoted by $f_1 [m] =1/(k_1l_{q1})$ and $f_2 [m]=1/(k_2l_{q2})$ and the drifts between the elements by 
$s_1$, $s_2$ and $s_3$. For simplicity, the center of consecutive dipoles is considered as the entrance and exit 
of the TME cell.

The horizontal emittance of the beam in an iso-magnetic ring:
\begin{equation}
\label{eq:emit}
\epsilon_x  =  \frac{C_q \gamma^2}{{\cal J}_x \rho_x} \langle {\cal H}_x \rangle,
\end{equation}
is determined by the  average dispersion invariant in the dipoles,
${\cal H}_x$ = $\gamma_x D_x^2 + 2 \alpha_x D_x D_x^\prime + \beta_x {D_x^\prime}^2$, 
where $\alpha_x,~\beta_x,~\gamma_x$ are the twiss parameters and $D_x$, $D_x^\prime$ the dispersion 
and its derivative. The parameter $C_q=3.84 \times 10^{-13}$~m is the quantum fluctuation coefficient for the 
electron, $\gamma$ the relativistic factor, ${\cal J}_x$  the damping partition number, and $\rho_x$ the bending radius. 
The minimization of the dispersion invariant average, provides the conditions  of $\beta_x$ and $D_x$ at the 
center of the dipole, for achieving  the theoretical minimum emittance~\cite{MinimumEmitRadiation}: 
\begin{eqnarray}
\label{eq:betaeta}
\beta_{x\mathrm{c}}^{\mathrm{min}}=\frac{l_d}{2\sqrt{15}}, ~~~ \alpha_{x\mathrm{c}}=0, ~~~D_{x\mathrm{c}}^{\mathrm{min}}=\frac{\theta l_d}{24}, ~~~D'_{x\mathrm{c}}=0,
\end{eqnarray}
where $\theta=\frac{l_d}{\rho_x}=\frac{2\pi}{N_d}$ is
the bending angle for $N_d$ dipoles in the ring. For a general TME cell, the geometrical emittance can be expressed as: 
\begin{equation}
\label{eq:emitgen}
	\epsilon_x = \frac{C_q \gamma^2}{{\cal J}_x \rho_x} \left[ \frac{1}{\beta_{x\mathrm{c}}} \left(D_{x\mathrm{c}}^2-\frac{\theta D_{x\mathrm{c}} l_d}{12}+\frac{\theta^2 l_d^2}{320}\right)+\frac{\theta^2 \beta_{x\mathrm{c}}}{12} \right],
	\end{equation}
where $D_{x\mathrm{c}}$ and $\beta_{x\mathrm{c}}$ the dispersion and beta functions at the center of the dipole. Substituting the values 
of $D_{x\mathrm{c}}^{\mathrm{min}}$ and $\beta_{x\mathrm{c}}^{\mathrm{min}}$ in Eq.~\eqref{eq:emitgen} with their TME expressions of 
Eq.~\eqref{eq:betaeta}, the emittance 
becomes $\epsilon_{x\mathrm{TME}}$=$\cal F$ $C_q\gamma^3\theta^3$. The scaling factor $\cal F$ for the TME lattice 
is $\cal F$=$\frac{1}{12\sqrt{15}{\cal J}_x}$ and the damping partition number ${\cal J}_x\approx1$, in the case of isomagnetic rings, based on 
dipoles without quadrupole gradient~\cite{ref:Lee}.  Defining the ratios $\beta_r=\frac{\beta_{x\mathrm{c}}}{\beta_{x\mathrm{c}}^{\mathrm{min}}}$ and  
$D_r=\frac{D_{x\mathrm{c}}}{D_{x\mathrm{c}}^{\mathrm{min}}}$, it is useful to define the emittance detuning factor~\cite{PhysRevSTAB.4.021001}: 
\begin{equation}
\label{eq:emitdet}
	\epsilon_r =  \frac{9 + 4 \beta_r^2 + 5 (D_r -2) D_r}{8 \beta_r},
\end{equation}
with  $\epsilon_x=\epsilon_r \cdot \epsilon_{x,\mathrm{TME}}$. The detuning factor is an indication of how much the emittance deviates 
from its theoretical minimum, for a given set of optics parameters at the center of the cell.

Inverting Eq.\eqref{eq:emitdet} and solving with respect to $\beta_r$, the following expression is computed: 
\begin{eqnarray}
	& \beta_r  = \epsilon_r \pm \frac{1}{2}\sqrt{-9 + 4 \epsilon_r^2 - 5 (-2 + D_r) D_r}.
\label{eq:betacd}
\end{eqnarray}
The quadratic dependence on $D_r$ of the argument in the square root in Eq.~\eqref{eq:betacd}, sets an upper 
and a lower limit for the dispersion at the center of the dipole, in order for $\beta_r$ to be a real number:
\begin{eqnarray}
	 1-\frac{2 \sqrt{-1+\epsilon_r^2}}{\sqrt{5}} & \leq D_r & \leq 1+\frac{2 \sqrt{-1+\epsilon_r^2}}{\sqrt{5}}.
\label{eq:etacd}
\end{eqnarray}

\subsection{Analytical solutions for the quadrupole strengths}

The beta $\beta_{x\mathrm{c}}$ and dispersion $D_{x\mathrm{c}}$ functions, at the dipole center, impose two independent optics constraints 
and thus at least two quadrupole families are needed for achieving them. The horizontal optics functions 
are fully controlled by these two pairs of quadrupoles, whereas in the absence of additional knobs, the vertical plane 
optics is also uniquely defined. Using basic linear optics arguments and the thin lens approximation 
and for specific $\beta_{x\mathrm{c}}$ and $D_{x\mathrm{c}}$ at the center of the dipole (or $\beta_r$ and $D_r$), analytical expressions 
can be derived for the strengths of the quadrupoles:
\begin{equation}
\label{eq:f1f2}
 \begin{split} 
	 f_1  = & \frac{s_2 (4 l_d s_1 + l_d^2 + 8 D_{x\mathrm{c}} \rho_x)}{(4 l_d s_1  + l_d^2 + 8 D_{x\mathrm{c}} \rho_x) + 4 l_d s_2  - 8 D_s \rho_x} = \frac{l_ds_2\left(12s_1+l_d(D_r+3)\right)}{12l_d(s_1+s_2)+l_d^2(D_r+3)-24 D_s\rho_x}, \\ 
	 f_2  = & -\frac{8 s_2 D_s \rho_x}{(4 l_d s_1 + l_d^2 + 8 D_{x\mathrm{c}} \rho_x) -  8 D_s \rho_x}=-\frac{24s_2 D_s\rho_x}{12l_ds_1+l_d^2(D_r+3)-24 D_s\rho_x},
\end{split}
\end{equation}
which are parametrized with the drift lengths  $s_1,~s_2,~s_3$. The parameter $D_s$ is the dispersion at the 
center of the cell (between two mirror symmetric quadrupoles) and is a function of the drift lengths, the optics 
functions at the dipole center and the bending characteristics: 

\begin{equation}
\label{eq:etas}
	D_s  = \frac{A \pm \sqrt{A^2+A B C}}{64 B \rho_x^2},
\end{equation}
where:
\begin{eqnarray}
\label{eq:etasABC}
	A &=&  8 s_2 \rho_x \left[ l_d^4 + 64 D_{x\mathrm{c}}^2 \rho_x^2 + 16 l_d^2 (\beta_{x\mathrm{c}}^2 - D_{x\mathrm{c}} \rho_x )\right] \nonumber \\
	  &=& \frac{8}{45}l_d^4 s_2 \left[12 \beta_r^2+5(D_r-3)^2\right]\rho_x, \nonumber \\
	B &=& l_d \left(2 s_1 l_d + l_d^2 +8 \beta_{x\mathrm{c}}^2\right)-8 \left(2 s_1+l_d\right) D_{x\mathrm{c}} \rho_x \nonumber \\
	  &=& \frac{1}{15}l_d^2\left[l_d(15+2 \beta_r^2-5 D_r)-10s_1(D_r-3)\right],  \nonumber \\
	C &=& \frac{16 s_3 \rho_x (4 s_1 l_d+l_d^2+8 D_{x\mathrm{c}} \rho_x)}{s_2} \nonumber \\
	  &=& \frac{16 l_d s_3\left[12 s_1+l_d(D_r+3)\right]\rho_x}{3 s_2}.
\end{eqnarray}
The calculation of $D_s$ springs from the symmetry requirement at the middle of the cell, $\alpha_x=0$. 
By applying the TME conditions at the middle of the dipole ($\alpha_x$=0, $D'_x$=0), 
the $\alpha_x$ function at the middle of the cell has a quadratic dependence on ($D_s^{-1}$), which results in the two solutions, with opposite sign in the second component, for $D_s$.

The horizontal and vertical phase advances of the cell can be defined through the trace of the cell 
transfer matrix
and from this, the horizontal phase advance can be written in a simple form as:
\begin{eqnarray}
\label{eq:cosphix}
 \cos \mu_x = & \dfrac{(l_d^2-8 D_{x\mathrm{c}} \rho_x)^2 - 16 l_d^2 \beta_{x\mathrm{c}}^2}{(l_d^2-8 D_{x\mathrm{c}} \rho_x)^2 + 16 l_d^2 \beta_{x\mathrm{c}}^2}= \dfrac{5(D_r-3)^2 -12 \beta_r^2}{5(D_r-3)^2 +12 \beta_r^2}.
\end{eqnarray}
For $D_r=\beta_r=1$, $\mu_x=\arccos(1/4)=284.5^o$ independent on any cell parameter, which is a known property of the TME cells~\cite{PhysRevSTAB.4.021001}. 
The expression for the vertical phase advance has a more complicated form: 
 \begin{equation}
 \label{eq:cosphiy2}
 \begin{split} 
      \cos \mu_y = & 1+\frac{L_c}{f_1}+\frac{L_c}{f_2}+\frac{s_{23}(L_c-2s_{23})}{f_1^2}+\frac{s_3(L_c-2s_3)}{f_2^2}+\frac{2s_{23}L_c-2(s_{23}^2+s_3^2)}{f_1f_2}\\
		   & +\frac{s_2(s_{23}+s_3)(L_c-2s_{23})}{f_1^2f_2}+\frac{2s_2s_3(L_c-s_{23}-s_3)}{f_1f_2^2}
		   +\frac{s_2^2s_3(L_c-2s_{23})}{f_1^2f_2^2},
 \end{split}
 \end{equation}
where $L_c=l_d+2(s_1+s_2+s_3)$ the cell length and $s_{23}=s_2+s_3$. Unlike the horizontal plane, the  vertical 
phase advance depends not only on the optics functions at the dipole center but also the cell geometry.

\subsection{Momentum compaction factor}
An analytical expression can also be derived for the momentum compaction factor of the cell, under 
the TME conditions ($D_x'=0$ at the center of the dipole), and can be written in the form:
\begin{eqnarray}
\label{eq:ac}
  \alpha_p = \left\langle \frac{D_x}{\rho_x}\right\rangle = \frac{1}{l_d} \int_0^{l_d} \frac{D_x(s)}{\rho_x}ds = \frac{7}{12}\theta^2+\frac{2D_c}{\rho_x}=\frac{\theta^2}{12}(D_r+7),
\end{eqnarray}
depending only on the dipole characteristics and in particular, quite strongly on the bending angle, which explains 
the trend that the momentum compaction factor is reduced, when the dipoles become shorter and/or weaker. The momentum 
compaction factor for the absolute minimum emittance ($D_r=1$) is:
\begin{eqnarray}
\label{eq:acmin}
  \alpha_p^{\mathrm{TME}} = \frac{2\theta^2}{3},
\end{eqnarray}
which depends only on the dipole bending angle.

\subsection{\label{sec:level2} Optics stability}
The stability criterion for both horizontal and vertical planes is:
\begin{equation}
\label{eq:stabilitycrit}
 \mathrm{Trace}(M_{x,y})=2|\cos\mu_{x,y}|<2,
\end{equation}
where $M_{x,y}$ is the transfer matrix of the cell and $\mu_{x,y}$ are the horizontal and vertical phase advances 
per cell, respectively. 
The latest ensures the optics stability and can be used for constraining the cell characteristics (focal and drift 
lengths).

\subsection{The absolute minimum emittance limit}

In the absolute minimum emittance limit, where $\beta_r=D_r=1$, the parametric equations for the quadrupole strengths are reduced to:
\begin{eqnarray}
\label{eq:f1f2TME}
\begin{split}	 
	 f_1^{\mathrm{TME}}  = & \frac{(l_d+3s_1)(3l_d+5s_1)s_2}{(l_d+3s_1)(3l_d+5s_1)+(7l_d+15s_1)s_2 \pm 2 \sqrt{l_d(l_d+3s_1)(3l_d+5s_1)s_3+l_d^2s_2^2}}, \\
	 f_2^{\mathrm{TME}}  = & \frac{2l_ds_2s_3}{l_d(s_2+2s_3) \pm \sqrt{l_d(l_d+3s_1)(3l_d+5s_1)s_3+l_d^2s_2^2}}.
\end{split}
\end{eqnarray}
Applying the requirement of opposite sign quadrupole strengths in the above equations, thus $f_1 \times f_2 < 0$, in order to assure optics stability 
in both planes, the case of (-) sign of Eq.~\eqref{eq:etas} can be ignored. This shows that the dispersion 
at the symmetry point of the cell can never become negative. For the (+) sign, the following constraints are derived:
$$s_3>\frac{\left[l_d+3(s_1+s_2)\right]\left[3l_d+5(s_1+s_2)\right]}{4l_d} ~~~\mathrm{and} ~~~ s_2\geq\frac{(l_d+3s_1)(3l_d+5s_1)}{4l_d}$$
or
$$s_2+s_3 < \frac{(l_d+3s_1)(3l_d+5s_1)}{4l_d}$$

It is interesting to study the behavior of Eqs.~\eqref{eq:f1f2TME} in the limit where the drift spaces lengths are going to zero.  
They are then reduced to:
\begin{equation}
\begin{split}
 \label{eq:f1TMElimits}
      (f_1^{\mathrm{TME}},~f_2^{\mathrm{TME}}) & \xrightarrow{s_1 \to 0} \left(\frac{3l_ds_2}{3l_d+7s_2 - 2\sqrt{s_2^2+3l_ds_3}},~\frac{2s_2s_3}{s_2+2s_3-\sqrt{s_2^2+3l_ds_3}} \right),\\
      (f_1^{\mathrm{TME}},~f_2^{\mathrm{TME}}) & \xrightarrow{s_2 \to 0} (0, 0), \\
      (f_1^{\mathrm{TME}},~f_2^{\mathrm{TME}}) & \xrightarrow{s_3 \to 0} \left(\frac{s_2(3l_d+5s_1)}{3l_d+5(s_1+s_2)},~\frac{4l_ds_2^2}{3l_d^2+14l_ds_1+15s_1^2-4l_ds_2}\right), 
\end{split}
\end{equation}
In the limits where $s_1 \to 0$ or $s_3 \to 0$ both $f_1$ and 
$f_2$ converge to specific values, depending on the dipole length and on the drift spaces lengths. Thus, realistic 
solutions exist even if the first quadrupole $Q_1$ is placed exactly after the dipole, without any space between them, or if the 
two $Q_2$ quadrupoles are merged to 1. 
In the limit where $s_2 \to 0$ both the focal lengths $f_1$ and $f_2$ go to zero or the quadrupole strengths to 
infinity. A good separation of the two quadrupoles is thus necessary in order to have a feasible TME cell. 
In the limit of the absolute minimum emittance and of $s_2 \to 0$, the $\cos \phi_y$ function 
goes also to infinity verifying that those solutions are optically unstable. 

\subsection{\label{sec:level2} Magnet technology constraints}

Even if satisfied, the stability criteria do not necessarily guarantee technologically feasible magnet strengths. 
The pole tip field of the quadrupoles and chromatic sextupoles is constrained by the maximum value allowed by the 
chosen magnet technology. In addition, the radius of the magnets' aperture should be greater than a minimum 
value, defined by beam and lattice properties. 

The quadrupole gradient (expressed in [T/m]) is defined as $g = k(B \rho_x)$,
where $k$ the quadrupole strength and $B\rho_x$ the magnetic rigidity. From the definition of the pole tip field: 
$B_q = R \frac{\partial B_y}{\partial x}|_{y=0} = R g$, the gradient is $g=\frac{B_q}{R}$, where $R$ is the 
quadrupole aperture radius. 
Considering a circular beam pipe, the minimum required aperture radius in order to accept all the particles of the incoming beam, 
for a non-Gaussian beam distribution, is defined 
by the displacement of the particles with the maximum action in the beam, defined by an emittance $\epsilon_{\mathrm{max}}$ 
and a momentum deviation $(\delta p/p_0)_{\mathrm{max}}$~\cite{ref:WolskiNLCPDR}:
\begin{equation}
\label{eq:rmin}
R_{\mathrm{min}} = \sqrt{2 \beta \epsilon_{\mathrm{max}}} + (\frac{\delta p}{p_0})_{\mathrm{max}} \cdot D+d_{\mathrm{co}},
\end{equation}
where $\beta$ and $D$ the beta and dispersion functions at this location,  $(\delta p/p_0)$ the total energy spread 
of the beam and $d_{\mathrm{co}}$ a constant reflecting the tube thickness, mechanical tolerances and maximum orbit distortion. 
For a Gaussian beam distribution, Eq.~\eqref{eq:rmin} becomes:
$R_{\mathrm{min}} = \sqrt{2 \beta \epsilon_{\mathrm{max}}+((\frac{\delta p}{p_0})_{\mathrm{max}}\cdot D)^2}+d_{\mathrm{co}}$.
The $R_{\mathrm{min}}$ can be computed for each element of the cell and takes its maximum value at the center of the
quadrupoles, where the beta functions become maximum. The magnet technology constraint for the quadrupole gradient or  
strength is then:
\begin{eqnarray}
	g \leq \frac{B_q^{\mathrm{max}}}{R_{\mathrm{min}}} ~~ \mathrm{or} ~~ \frac{1}{f l_q} = k \leq \frac{1}{(B\rho_x)} \frac{B_q^{\mathrm{max}}}{R_{\mathrm{min}}}\;.
\label{eq:gcon}
\end{eqnarray}

In a similar way, a magnet technology constraint can be set for the sextupole strengths. As already mentioned, 
the TME cells are intrinsically high chromaticity cells when targeting to their theoretical minimum emittance limit, 
as low dispersion and strong focusing are needed to achieve the ultra low emittance. 
The high chromaticity 
requires strong sextupoles for the chromaticity correction, reducing the dynamic aperture of the machine.
The sextupoles used for the natural chromaticity correction are usually placed close to the quadrupoles, to large 
dispersion and beta function regions. In order to simplify the calculations, the sextupoles are considered to be 
placed on top of the quadrupoles, with equal lengths. The pole-tip field for the sextupoles is 
$B_s=(B \rho_x) b_2 R^2=\frac{1}{2}R^2 \frac{\partial^2 B_y}{\partial x^2}|_{y=0}$ and the sextupole gradient 
$(B \rho_x) b_2 = B_s/R^2$. As the sextupoles are set to cancel the chromaticity induced by the quadrupoles, the sextupole 
strengths can be calculated by:
\begin{eqnarray}
	\xi_x & = & -\frac{1}{4 \pi} \oint{\beta_x[K_x(s)-S(s)D(s)]ds}=0, \nonumber\\
	\xi_y & = & -\frac{1}{4 \pi} \oint{\beta_y[K_y(s)+S(s)D(s)]ds}=0,
\label{eq:chromtot}
\end{eqnarray}
where $K_{x,y}$ the focusing and defocusing quadrupole strengths and $S=\frac{b_2}{(B\rho_x)}$ the sextupole strengths. 
Evaluating the above integrals along the cell, the expressions for the sextupole strengths are:
\begin{eqnarray}
	S_1 & = &-\frac{2 \xi^q_y \pi \beta_{x\mathrm{,d}}+2 \xi^q_x \pi \beta_{y\mathrm{,d}}}{l_q \beta_{x\mathrm{,f}} \beta_{y\mathrm{,d}} D_{x\mathrm{,f}}-l_q \beta_{x\mathrm{,d}}\beta_{y\mathrm{,f}} D_{x\mathrm{,f}}}, \nonumber\\
	S_2 & = & \frac{2 \xi^q_y \pi \beta_{x\mathrm{,f}}+2 \xi^q_x \pi \beta_{y\mathrm{,f}}}{l_q \beta_{x\mathrm{,f}} \beta_{y\mathrm{,d}} D_{x\mathrm{,d}}-l_q \beta_{x,d}\beta_{y,f} \eta_{x,d}},
\label{eq:chromq}
\end{eqnarray}
where $\xi^q_{x,y}= -\frac{1}{4 \pi} \oint{\beta_{x,y} K_{x,y} ds}$ and $l_q$ the length of the quadrupoles. 
For simplicity, we consider all the quadrupoles to have the same length. In the expressions above, the index $f$ denotes 
the values of the optics functions on the focusing quadrupoles while $d$ the values on the defocusing quadrupoles.
In order to have feasible solutions, these values need to satisfy the constrain:
\begin{eqnarray}
	S \leq \frac{B_s^{\mathrm{max}}}{R_{\mathrm{min}}^2} \frac{1}{(B\rho_x)^2}.
\label{eq:Sconst}
\end{eqnarray}

Equations \eqref{eq:betacd}, \eqref{eq:etacd}, \eqref{eq:f1f2}, \eqref{eq:cosphix}, \eqref{eq:gcon}, \eqref{eq:Sconst} 
fully describe the linear optics of the TME cell. The parameter space of the cell, including geometrical and optical properties, can be 
determined giving the possibility to optimize the cell according to any design requirements.

\section{\label{sec:NumApplication}Numerical Application}

The analytical parameterization can be used to study the performance of any TME cell 
of interest. Some numerical examples, applicable to the CLIC Pre-damping rings (PDR) lattice 
design, will be used to demonstrate the results. The energy of the CLIC Damping Rings complex of 2.86~GeV~\cite{ref:CDR-DR} 
and a dipole field of 1.2~T are used. The required output normalized emittance from the CLIC PDR is 63~$\mu$m-rad. 
Leaving a blow up margin of 10~\%, and using Eq.~\eqref{eq:emitgen}, at least 19 dipoles (or TME cells) 
of 1.2~T field and $\theta=2\pi/N_d \approx 19^o$ 
bending angle, are needed. The example of a TME cell with 38 dipoles of 1.2~T bending field and $\theta \approx 9.5^o$ 
bending angle is also discussed. 
In order to set the feasibility constraints of the quadrupole and sextupole magnets, the maximum pole-tip field of the 
quadrupoles is set to $B_q^{\mathrm{max}}=1.1$~T and for the sextupoles $B_s^{\mathrm{max}}=0.8$~T, which are typical values 
for normal-conducting magnets. Both quadrupole and sextupole lengths are set to $l_q=0.3$~m. 
Fixing those parameters, the free parameters left are the drift space lengths, $s_1$, $s_2$ and $s_3$, and the 
emittance $\epsilon_x$, or the detuning factor $\epsilon_r$. The parametrization with respect 
to drift spaces lengths and with respect to the emittance are treated separately.

\subsection{\label{sec:DriftParam}Parametrization with the drift lengths}

\begin{figure}[ht]
   \centering   
   \includegraphics*[width=0.47\textwidth,trim=3cm 8cm 3cm 8cm]{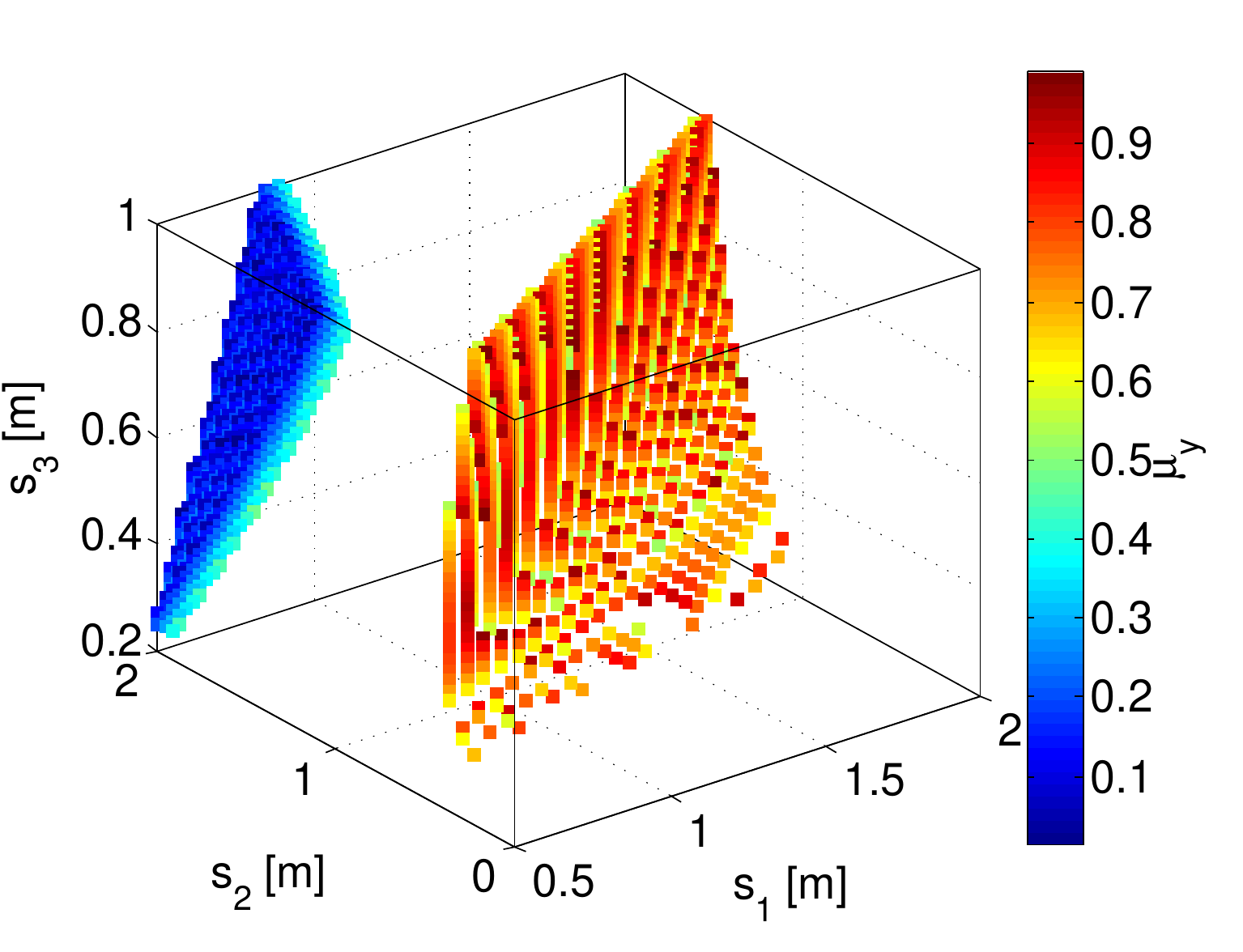}
   \includegraphics*[width=0.49\textwidth,trim=3cm 8cm 3cm 8cm]{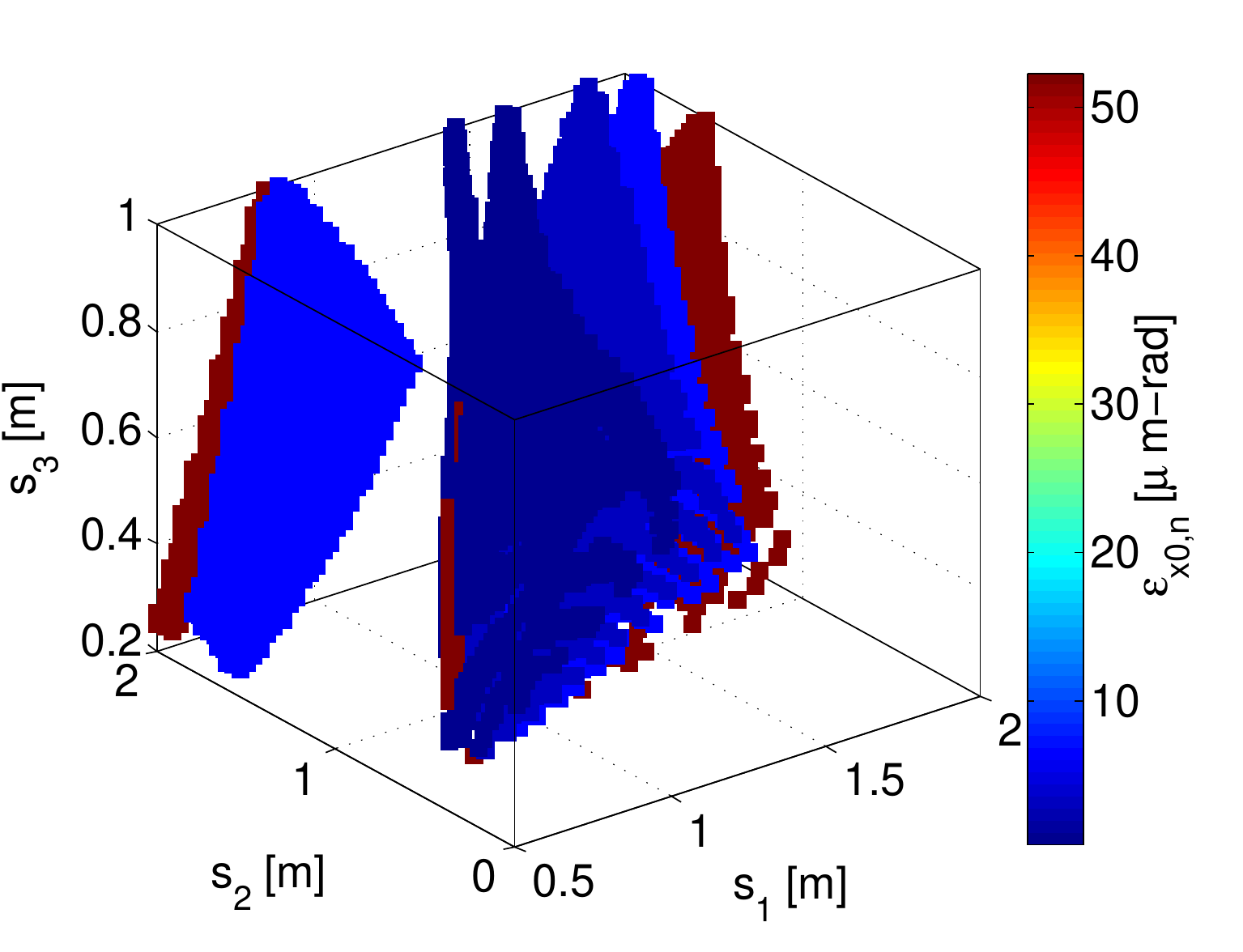}   
   \caption{Left: Parametrization of the vertical phase advance of a TME cell with a dipole bending angle 
	  of $\theta=2 \pi/19$ with the drift spaces lengths $s_1$, $s_2$, $s_3$, when targeting the 
	  theoretical minimum emittance. Only solutions providing optical stability are presented. Right: 
	  Dependence of the stability region on the dipole's bending angle or the theoretical minimum emittance.}
   \label{fig:stability}
\end{figure}
At first, a constant emittance is considered and we seek the drift spaces lengths 
satisfying the stability constraints in both horizontal and vertical planes. 
By construction the horizontal plane is always optically stable, thus this constrain comes solely from the vertical plane. 
The vertical phase advance (defined in Eq.~\eqref{eq:cosphiy2}) was calculated for all combinations of 
$s_1$, $s_2$ and $s_3$, for $s_1\in (0.5, 2)$~m, $s_2\in (0.5, 2)$~m and $s_3\in (0.25, 1)$~m.  
The optically stable solutions for a TME cell with bending angle of $\theta=2\pi/19$ , corresponding to a
dipole length of $l_d$=2.6~m and an equilibrium emittance of $\epsilon_x^{\mathrm{TME}}$=~52~$\mu$m-rad, parametrized 
with $s_1$, $s_2$ and $s_3$ are shown in the left part of Fig.~\ref{fig:stability}. There are two manifolds of 
stable solutions, clearly distinguished with respect to the vertical phase advance. Solutions with small $s_1$ and 
moderate to large $s_2$ correspond to small phase advances ($\mu_y \textless 0.5$), while small to moderate $s_2$ 
solutions correspond to large phase advances ($\mu_y \textgreater 0.5$). The optics stability is independent 
on the value of $s_3$, as there are stable solutions for each value of $s_3$, for both regions of solutions. 
The right part of Fig.~\ref{fig:stability} shows how the stability regions are changing for different dipole 
bending angles (or theoretical minimum emittances). Targeting to lower TME values, only the high phase 
advance solutions (small to moderate $s_2$) provide optical stability.

\begin{figure}[pht]
    \centering    
    \includegraphics*[width=0.4\textwidth]{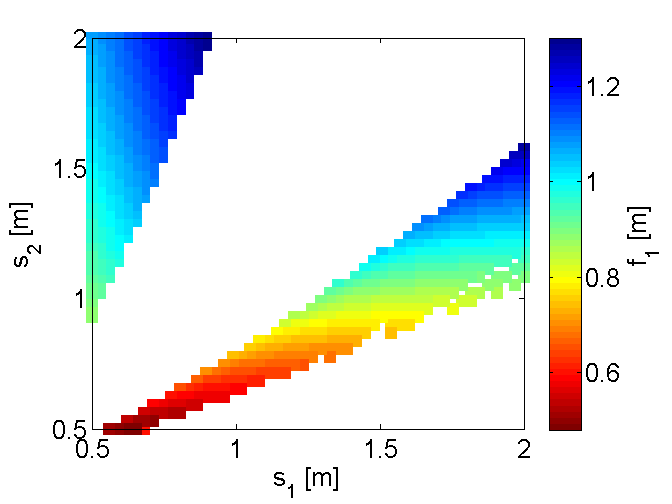}
    \includegraphics*[width=0.4\textwidth]{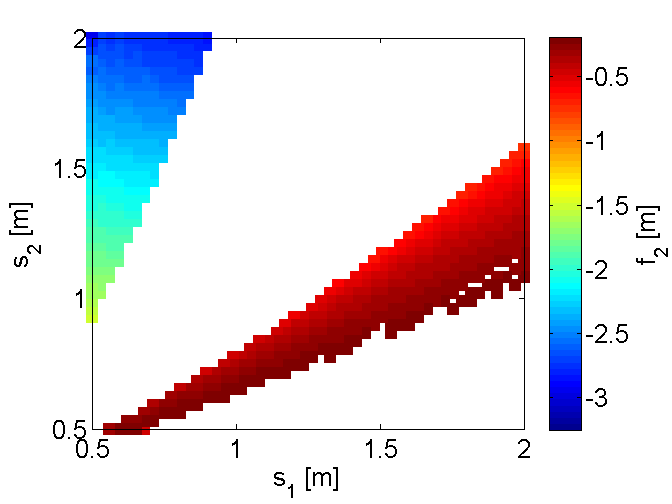}
    \includegraphics*[width=0.4\textwidth]{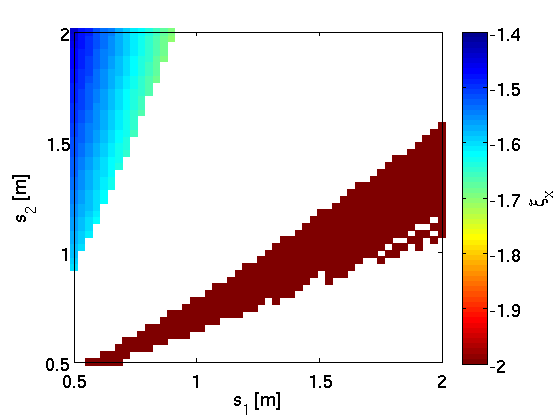}
    \includegraphics*[width=0.4\textwidth]{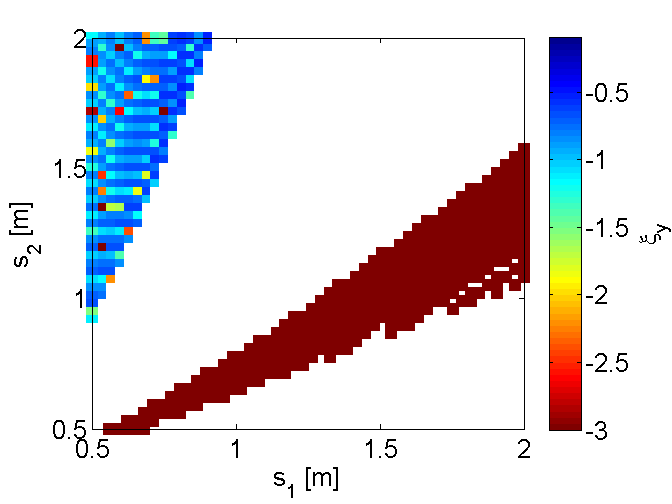}
    \caption{Parameterization of the quadrupole focal lengths (top), $f_1$ (left) and $f_2$ (right) and 
    the horizontal (bottom, left) and vertical (bottom, right) chromaticities with the drift spaces 
    lengths, $s_1$, $s_2$ providing optical stability.}
    \label{fig:s1s2s3f1f2}
\end{figure}
The combinations of drift lengths satisfying the stability requirements are then applied to Eqs.~\eqref{eq:f1f2}-\eqref{eq:cosphiy2},  
for the calculation of all cell properties.
Fig.~\ref{fig:s1s2s3f1f2} shows the parametrization of the quadrupole focal lengths (top), $f_1$ (left) and $f_2$ (right) 
and the horizontal (bottom, left) and vertical (bottom, right) chromaticities with the drift spaces lengths, $s_1$, $s_2$.
The two manifold of stable solutions are clearly distinguished with respect to the quadrupole focal 
lengths (especially for the defocusing quad) and the horizontal and vertical chromaticities. The 
small $s_1$ - large $s_2$ region (low vertical phase advance solutions) correspond to weaker quadrupole strengths 
and smaller chromaticities. 
The small to moderate $s_2$ region (large vertical phase advance solutions) correspond to strong quadrupole focal 
lengths (especially the vertical one) and large chromaticities. If magnet feasibility constraints are applied to these solutions, 
the latest region is rejected. 
The above observations can lead to the conclusion that proper choice of the drift spaces lengths triplet ($s_1$, $s_2$, $s_3$) can assure the stability 
of the motion, and lead to the minimization of the quadrupole strengths (maximum focal lengths) and to the 
minimization of the cell chromaticities in both planes, achieving always the same minimum emittance. 

\begin{figure}[ht]
    \centering
    \includegraphics*[width=0.45\textwidth,trim=3cm 8cm 2cm 8cm]{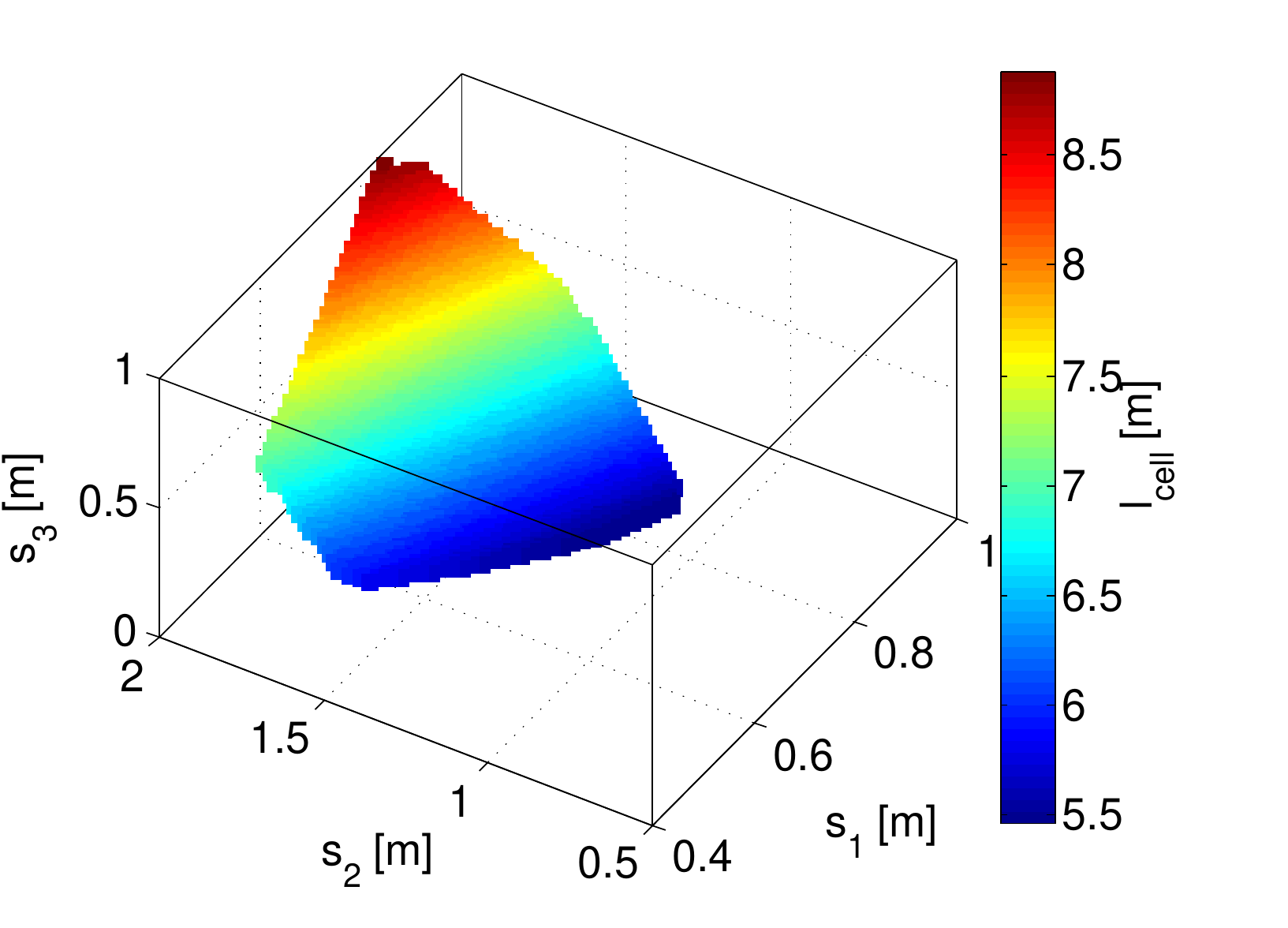}
    \includegraphics*[width=0.45\textwidth,trim=3cm 8cm 2cm 8cm]{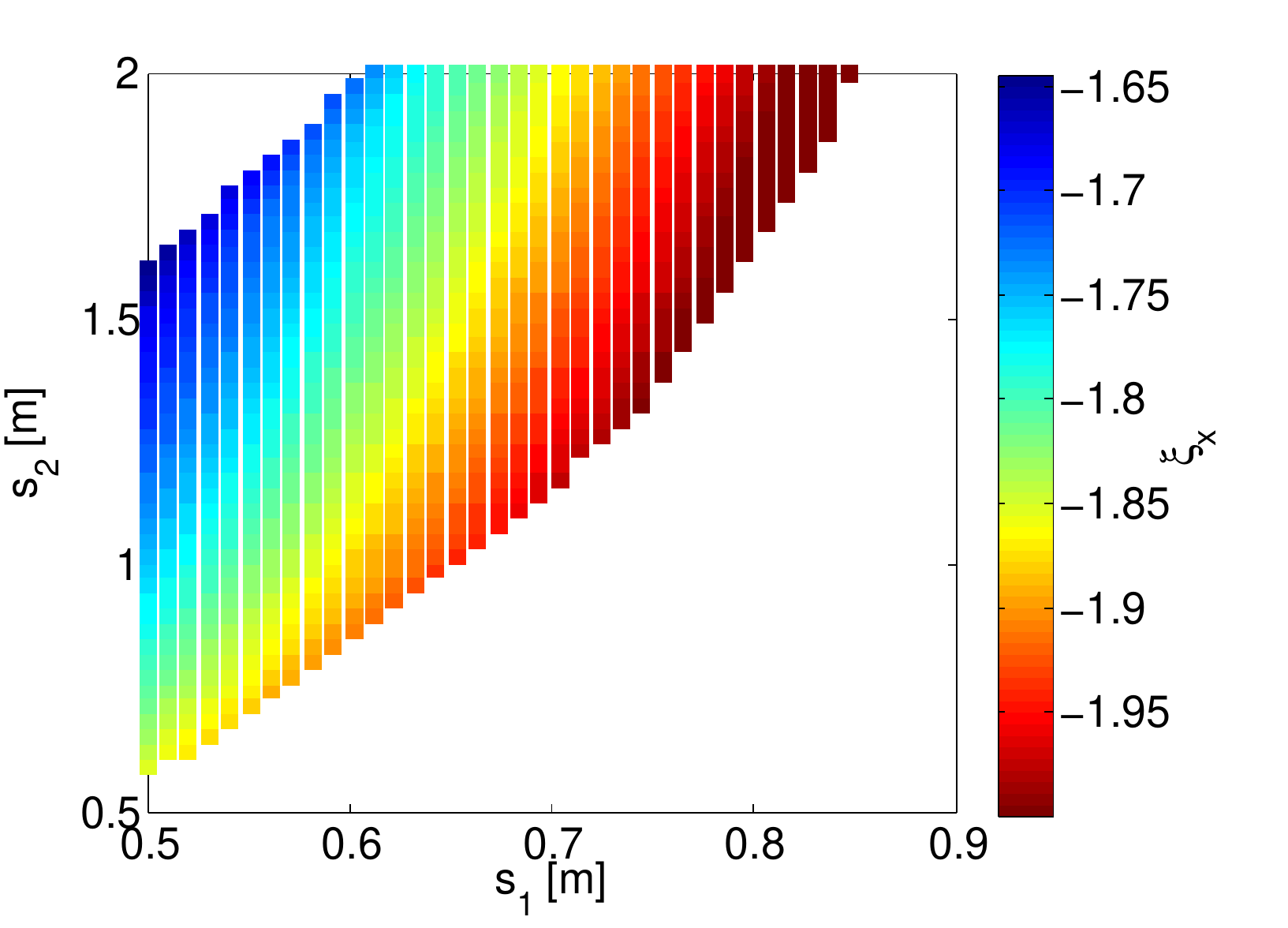}
    \caption{Left: The ($s_1$, $s_2$, $s_3$) combinations that provide the lowest chromaticity in 
    both planes ($\xi_{x,y} \leq -2$), colorcoded with the total length of the cell. Right: Projection of the low chromaticity solutions to the 
    ($s_1$, $s_2$) plane.}
    \label{fig:s1s2s3lowchrom}
\end{figure}
In the case of the low emittance rings lattice design, low chromaticity solutions are of interest for larger  
dynamic aperture. The cell length, on the other hand, is preferred to stay as compact as possible to minimize 
the circumference of the ring.
Fig.~\ref{fig:s1s2s3lowchrom} shows the ($s_1$, $s_2$, $s_3$) triplets 
for which the absolute chromaticity in both planes is less than 2, color-coded with the total cell length $\mathrm{l_{cell}}$ (left). 
In the right part of the figure, the projection of the solutions to the ($s_1$, $s_2$) plane is shown, 
color-coded with the horizontal chromaticity. In order to keep the chromaticity in low levels and the cell length as compact 
as possible, small values of $s_1$ and small to moderate values of 
$s_2$ are needed.
However, even the minimum possible chromaticity of this cell is quite large ($\xi_x\sim-1.65$).

\subsection{\label{sec:EmitParam}Parametrization with the emittance}

Having the drift lengths fixed, Eq.~\eqref{eq:f1f2} combined with Eqs.~\eqref{eq:betacd} and~\eqref{eq:etacd} are 
studied numerically for different detuning factors $\epsilon_r$. In this example, the dipole bending angle is set to 
$\theta=2\pi/38$ and the drift lengths to $s_1$=0.9~m, $s_2$=0.6~m and $s_3$=0.5~m. This configuration was found 
to be the optimal one for the CLIC PDR lattice design, as will be shown later. 

\begin{figure}[ht]
   \centering
    \includegraphics*[width=0.49\textwidth,trim=1.2cm 6cm 1.2cm 6cm]{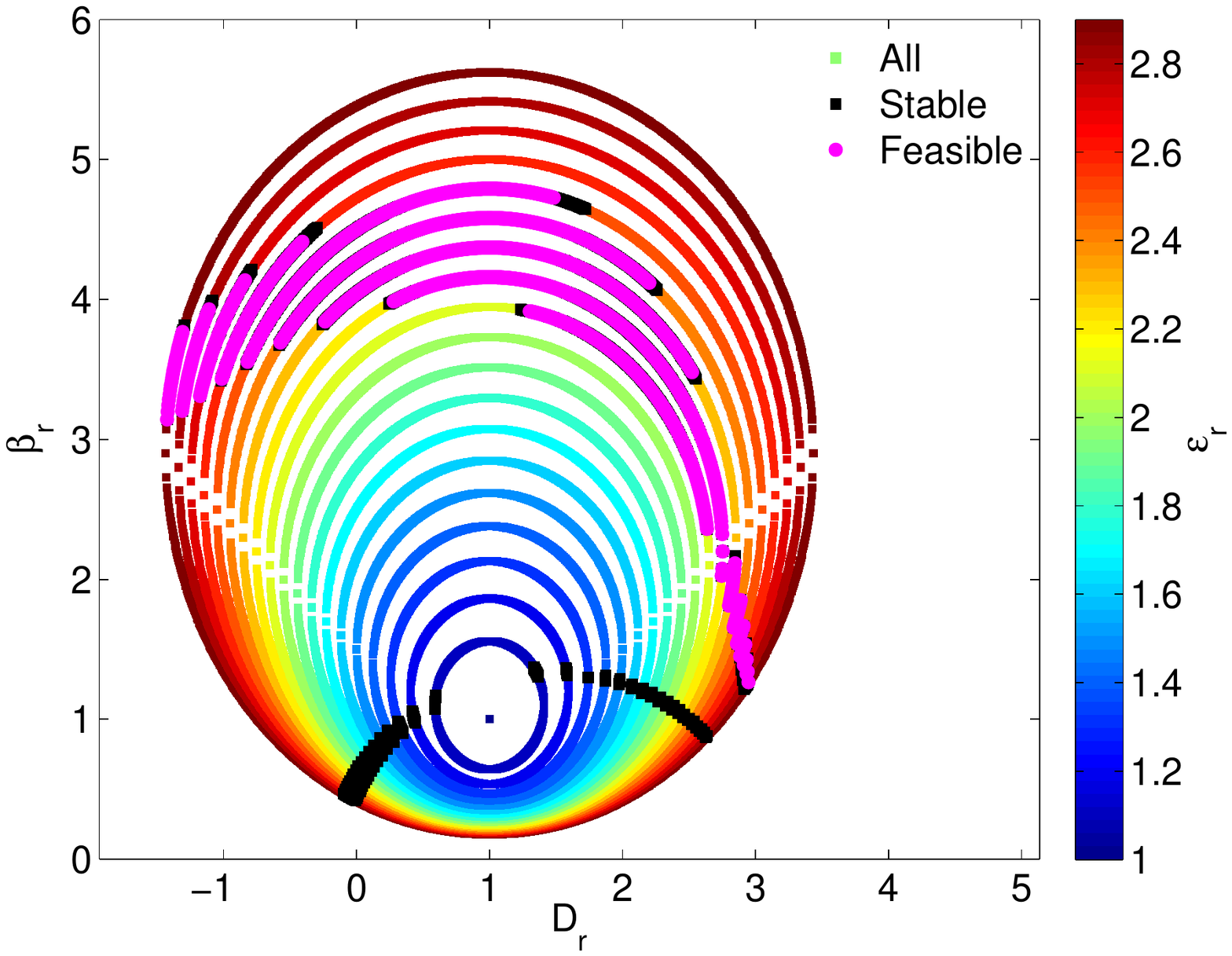}   
    \includegraphics*[width=0.49\textwidth,trim=1.2cm 6cm 1.2cm 6cm]{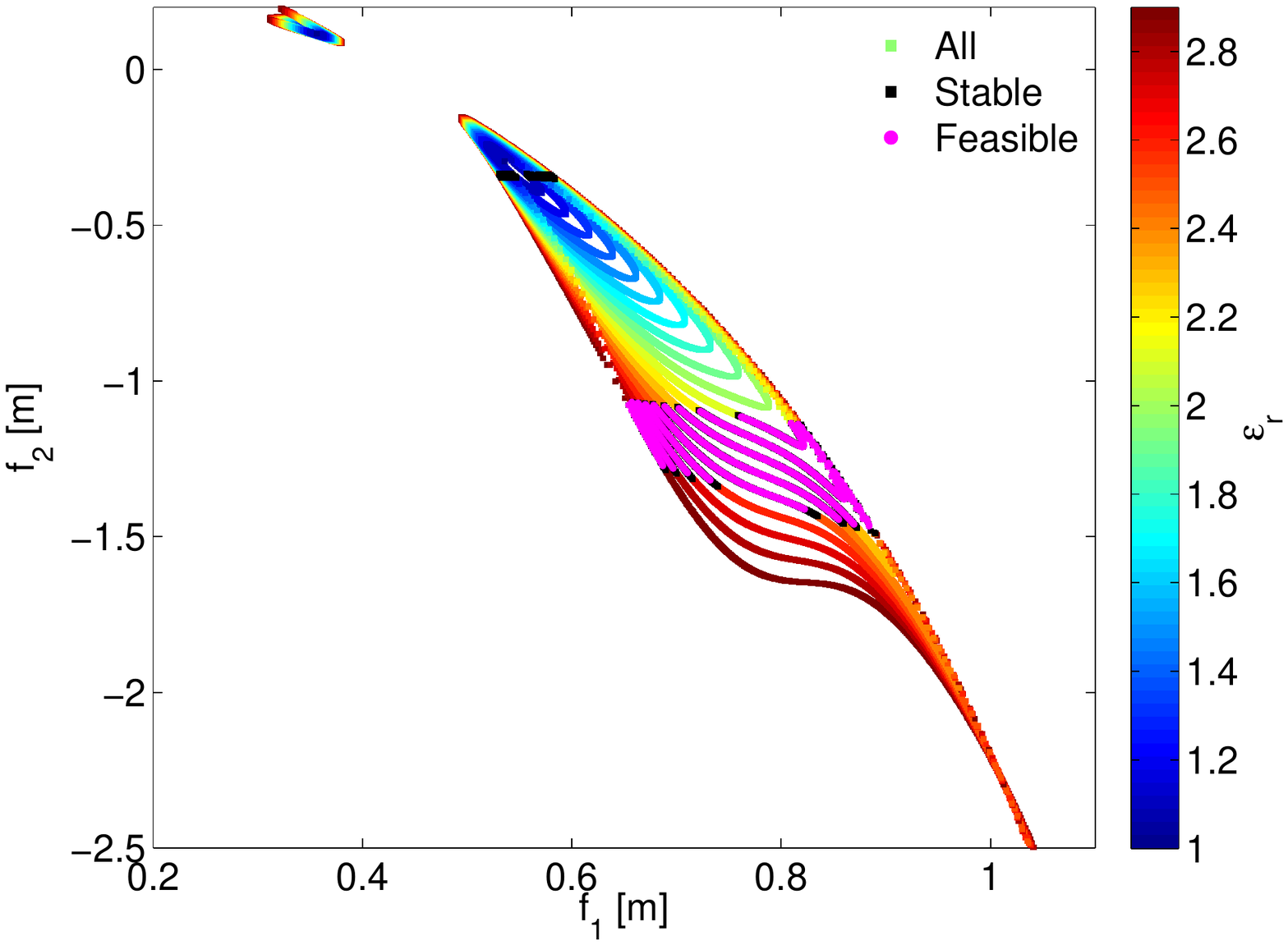}   
\caption{Parametrization of the relative horizontal beta $\beta_r$ and relative dispersion $D_r$ at the center of 
  the dipole (left) and the quadrupole focal lengths (right), with the cell detuning factor $\epsilon_r$. The stable solutions are 
  indicated with the black squares, while the stable and feasible ones with the magenta triangles.}
   \label{fig:emitpar}
\end{figure}
In order to achieve the absolute minimum emittance, only one pair of initial 
optics functions $(D_{x\mathrm{c}},~\beta_{x\mathrm{c}})$ or 
$(D_{r},~\beta_{r})$ exists~\cite{PhysRevSTAB.4.021001}. However, relaxing this requirement and 
detuning the cell to higher emittance values ($\epsilon_r > 1$), several pairs of $(D_{x\mathrm{c}},~\beta_{x\mathrm{c}})$ lying in 
elliptical curves can achieve the same emittance, as shown by equation~\eqref{eq:emitgen}. 
Fig.~\ref{fig:emitpar} (left) shows the solutions of $(D_{r},~\beta_{r})$ color-coded with the 
detuning factor $\epsilon_r$. Even though, by definition, all solutions are stable in the horizontal plane, 
only a small fraction of them satisfy the stability criteria of the vertical plane (black squares).
The parametrization of the focusing strengths with the emittance 
is displayed in  Fig.~\ref{fig:emitpar} (right), with the same color-convention as before. 
The $f_1$, $f_2$ pairs for the same detuning factor lie in distorted ellipses, which get more distorted  
while moving to high detuning factors. 
In order to tune the cell to the very low emittance, strong quadrupole strengths are needed and only one combination 
of ($f_1$, $f_2$) can tune the cell to the absolute emittance minimum. Moving away from the minimum emittance regime, 
the quadrupole strengths are relaxed for detuning factors greater than 2. In the upper left corner of the plot, 
solutions with both $f_1$ and $f_2$ positive, cannot provide stability as they always provide defocusing in the vertical plane. 
It is interesting to notice, that by changing 
the values of $f_1$ and/or $f_2$ by a small amount, the system remains stable if tuned in the relaxed $\epsilon_r$ 
regime but can easily get unstable if tuned to the absolute minimum emittance conditions.

\begin{figure}[ht]
  \centering
  \includegraphics*[width=0.32\textwidth]{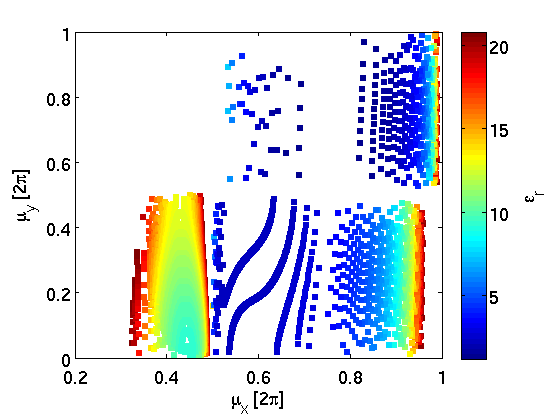}
  \includegraphics*[width=0.32\textwidth]{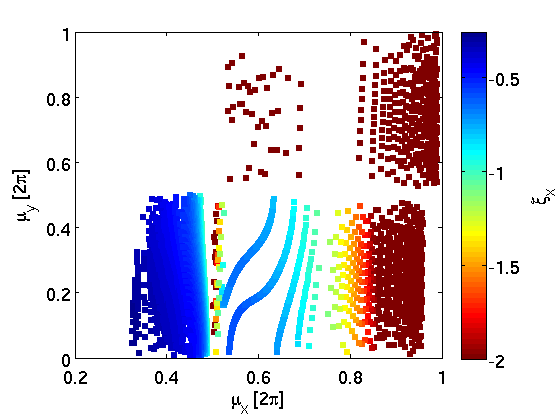}
  \includegraphics*[width=0.32\textwidth]{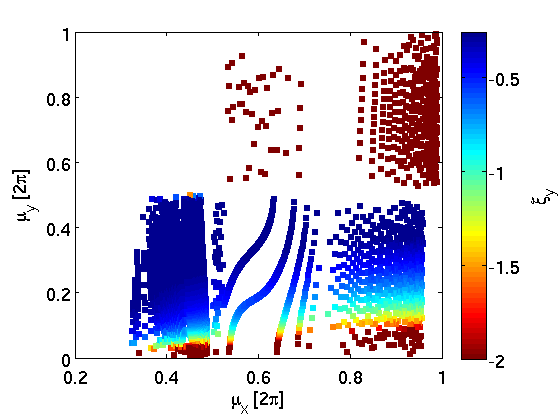}
  \includegraphics*[width=0.32\textwidth]{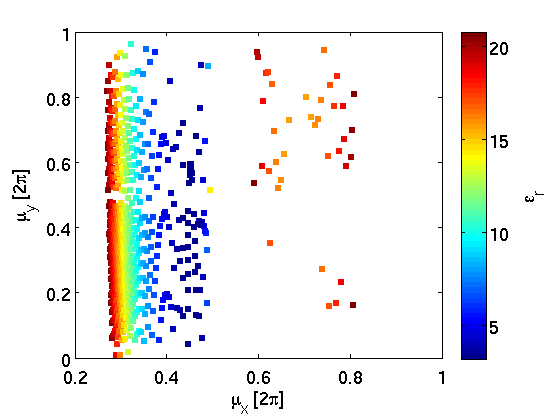}
  \includegraphics*[width=0.32\textwidth]{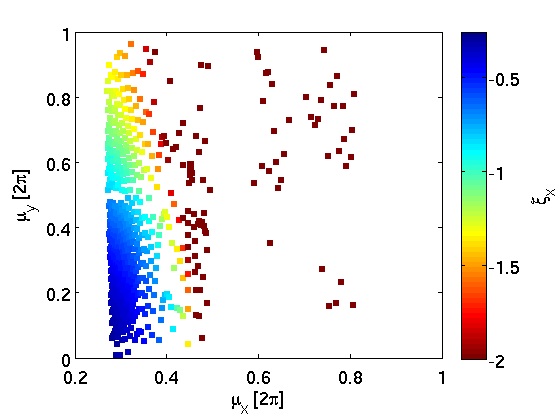}
  \includegraphics*[width=0.32\textwidth]{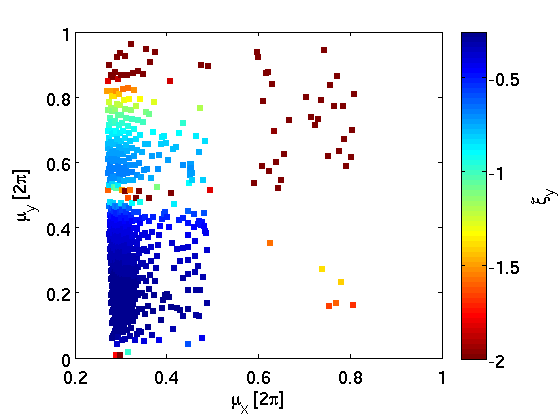}
   \caption{Parametrization of the cell detuning factor $\epsilon_r$ (left), and the horizontal (middle) and vertical (right) chromaticities 
	    with the horizontal and vertical phase advances of the cell is presented for each case, for a conventional (top) and a modified (bottom) 
	    TME cell.}
   \label{fig:muxVsmuyVsDetFact}
\end{figure}

Scanning in a broader range of the detuning factor, two different types of solutions survive the stability criteria. Solutions with focusing 
$Q_1$ and defocusing $Q_2$ are presented in the top part of Fig.~\ref{fig:muxVsmuyVsDetFact}, while the opposite case is presented in the bottom. 
Following the convention of~\cite{ref:PhysRevSTAB.14.054002}, we will refer to the former case as conventional TME cell while to the later as 
modified TME cell.
The parametrization of the cell detuning factor $\epsilon_r$ (left), and the horizontal (middle) and vertical (right) chromaticities 
with the horizontal and vertical phase advances of the cell is presented for each case.
For a conventional TME cell the chromaticities get minimized in both planes towards small phase advances, while the emittance detuning 
factor gets large values. Large phase advances correspond to high chromaticity values and small detuning factors. 
It is interesting 
to notice that the high detuning factor solutions at large horizontal phase advances produce large chromaticities, as they 
correspond to minimum dispersion and beta functions at the center of the dipole which require strong focusing.
In the case of the modified TME cell, the chromaticities are minimized for small phase advances as well, however in this case solutions with small 
detuning factors also exist. This type of cells is discussed in detail in~\cite{ref:PhysRevSTAB.14.054002}.

\section{\label{sec:MADXComp}Validation of the method}

\begin{figure}[ht]
 \begin{center}
     \includegraphics*[width=0.32\textwidth]{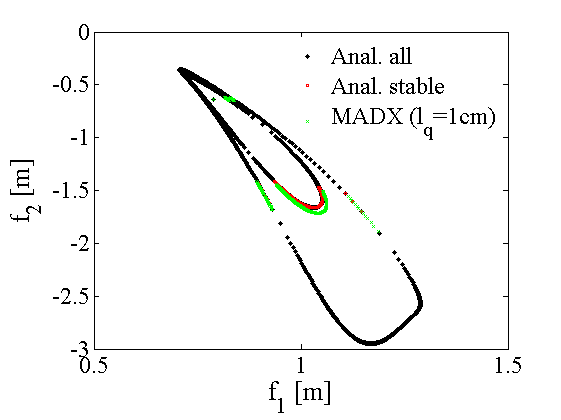} 
     \includegraphics*[width=0.32\textwidth]{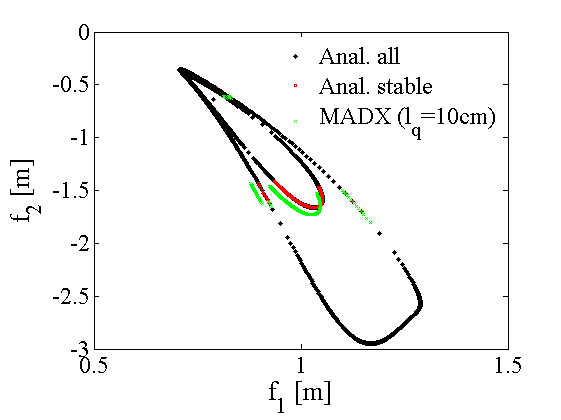} 
     \includegraphics*[width=0.32\textwidth]{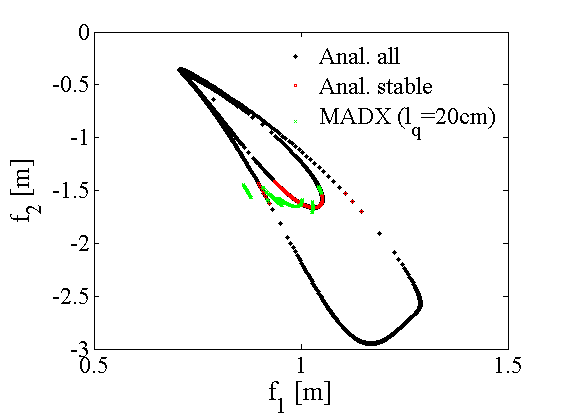} 
 \end{center}
\caption{Comparison between the analytical solution, and MADX simulations. 
 The analytical solutions are presented in black, the solutions satisfying the stability requirements in red 
 and the results from MADX for different quadrupole lengths: $l_q$=1~cm (top, left), $l_q$=10~cm (top, middle) 
 and $l_q$=20~cm (bottom), in green.}
\label{fig:cmpAnalMADX}
\end{figure}
The results of the analytical solution were compared to numerical simulations with 
MADX~\cite{bib:MADX} for the thin and thick lens cases. The three plots of Fig.~\ref{fig:cmpAnalMADX} show this comparison 
for three different values of the quadrupole lengths, $l_q=$1, 10 and 20~cm. 
The three curves of each plot in Fig.~\ref{fig:cmpAnalMADX}, represent three different detuning factors, $\epsilon_r$=1, 1.5  and 2. 
The analytical solutions are shown in black, the solutions satisfying the stability 
criteria  in red while the MADX solutions are presented in green. 
The agreement for the thin lens is excellent, demonstrating the validity of the analytical calculations.
It is very interesting that, even in the thick lens case, the agreement is still very good.
The analytical solution can be a very good approximation of 
the simulation results and can be helpful for the lattice optimization and understanding. In this way the optimal 
dipole characteristics,  the geometrical characteristics of the cell and the interesting phase advances can be defined.
It can also be very useful, for the definition of initial conditions to be used for the lattice design using numerical tools, 
whose optimization algorithms depend heavily on the initial values. 

This approach was thus used in order to define the optimal configuration and working point of the TME arc cells of the CLIC PDR lattice design.

\section{Application to the CLIC PDR design}
\label{sec:PDRDesign}

The CLIC Pre-damping rings provide the first stage of damping of the $e^+/e^-$ beams of the linear collider. 
They have to accommodate a 2.86~GeV beam with a large input emittance of 7~mm-rad, 
for positrons~\cite{ref:positronSource}, and damp it down to a normalized emittance of 
63~$\mathrm{\mu m}$-rad for injection into the main DR. The required input and output parameters are given in 
Table~\ref{tab:PDRParams}, for both electrons and positrons~\cite{ref:Params}. 
\begin{table}[!h]
 \caption{Parameters before the injection to the pre-damping rings and before the injection to the main damping rings.}
 \label{tab:PDRParams}
 \centering
 \begin{tabular}{lccc}\hline\hline
 \multirow{2}{*}{Parameters}		& \multicolumn{2}{c}{Injected}	& Extracted \\
 					&  $e^-$     	& $e^+$ 	& $e^-/e^+$ \\\hline
 Bunch Population~[$10^9$]		& 4.7		& 6.4		& 4.4		\\
 Bunch spacing~[ns]			& 0.5/1		& 0.5/1		& 0.5		\\
 Bunches/train				& 312/156	& 312/156	& 312/156		\\
 Number of trains			& 1/2		& 1/2		& 1/2		\\
 Repetition rate~[Hz]			& 50		& 50		& 50		\\
 Norm. horiz. emittance~[$\mu$ m-rad]	& 100 	& 7 x $10^3$	& 63 		\\
 Norm. vert. emittance~[$\mu$ m-rad]	& 100 	& 7 x $10^3$	& 1.5		\\
 Norm. long. emittance~[keV-m]	& 2.86		& 2288		& 143		\\\hline\hline
 \end{tabular}
\end{table}

Unlike the DR, the PDR lattice design is not driven by the emittance requirements~\cite{ref:CDR-DR}. The large energy spread and beam size of the 
injected beam, especially the one coming from the positron source, impose the requirements of large momentum acceptance and dynamic aperture. 
Thus the PDR 
lattice design is focused on the dynamic aperture optimization, providing at the same time a large enough momentum acceptance and the required output emittance.
Due to the more difficult characteristics of the positron beam, emphasis is given to the design of the positron pre-damping ring.
The rings were chosen to have a 
racetrack configuration with two arc sections and two long straight sections (LSS),
as a racetrack shape is the most compact one if only 2 dispersion free regions are required, which is valid for the case 
of the CLIC PDRs.
The arc sections are composed by TME cells, being the most compact low emittance cells. 
On the other hand, the LSS are composed by FODO cells filled with damping wigglers~\cite{ref:CDR-DR}. The damping wigglers are necessary 
to achieve the low emittance within a fast damping time, in order to fit into the 50~Hz repetition rate of the collider.
Permanent magnet wigglers of $B_w$=1.9~T peak field and $\lambda_w$=5~cm period provide the fast damping time and 
the required output emittance~\cite{ref:wiggler-des}.

\begin{figure}[ht]
   \centering
   \includegraphics*[width=0.45\textwidth,trim=2cm 8cm 2cm 8cm]{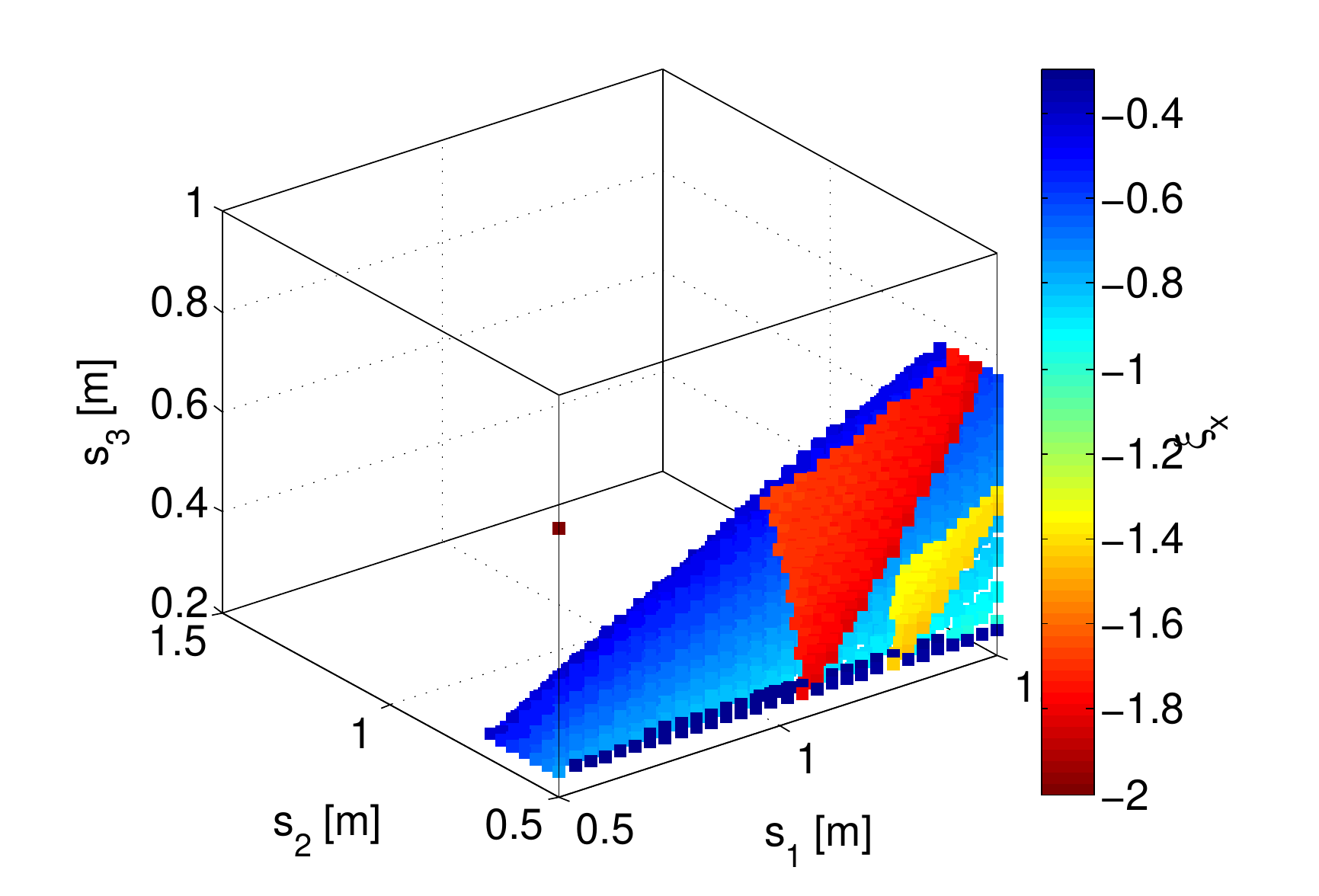}
   \includegraphics*[width=0.45\textwidth,trim=2cm 8cm 2cm 8cm]{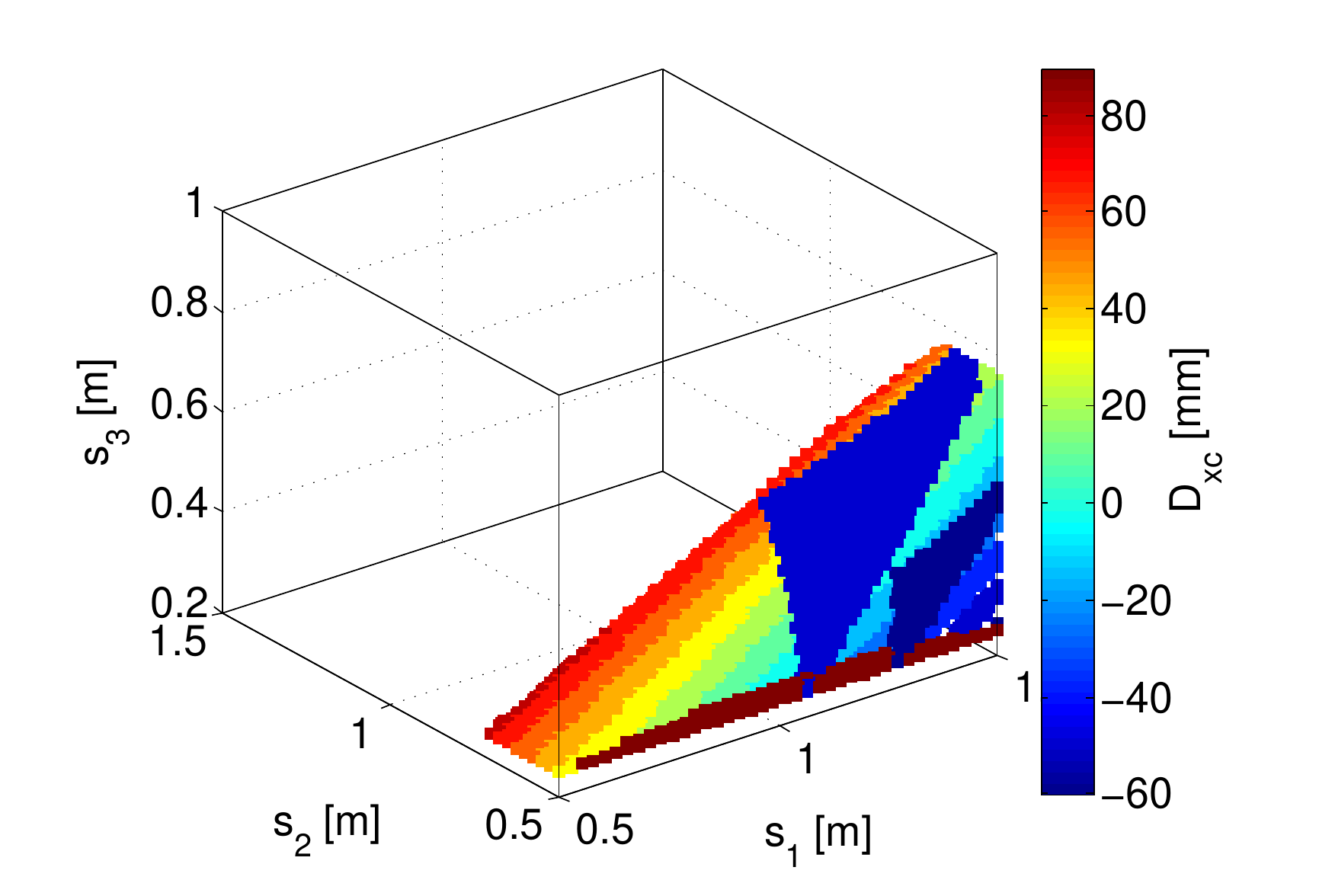}
   \includegraphics*[width=0.45\textwidth,trim=2cm 8cm 2cm 8cm]{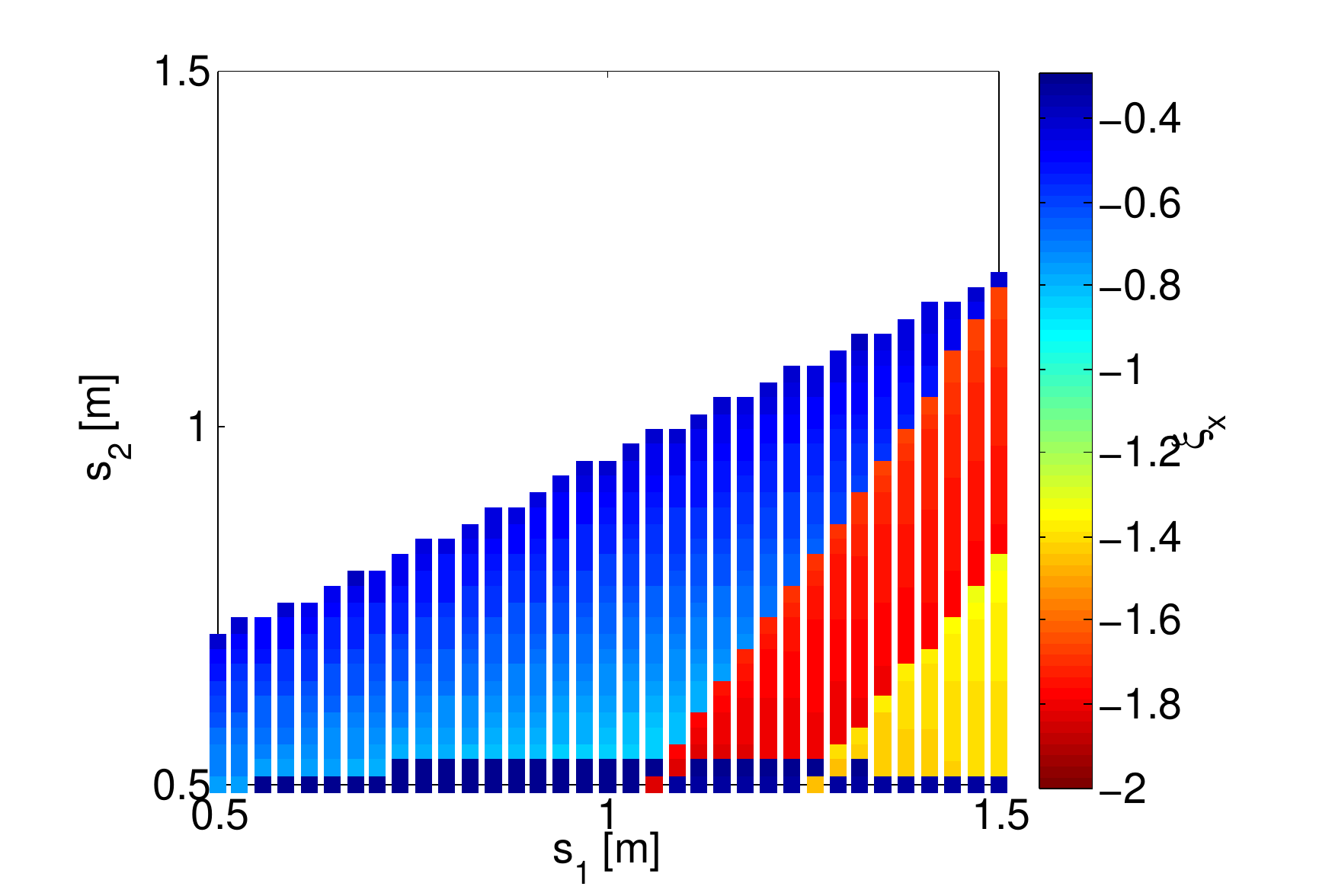}
   \caption{Top: Parametrization of the horizontal chromaticity (left) and the horizontal dispersion at the middle of 
	    the dipole (right), for a detuned cell ($\epsilon_r$=10), with the drift spaces lengths, $s_1$, $s_2$ and $s_3$. 
	    Bottom: Projection of the low chromaticity solutions in the ($s_1$, $s_2$) plane, color-coded with the horizontal chromaticity.}
   \label{fig:driftscandetuned}
\end{figure}
For the optimization of the CLIC PDR lattice design low chromaticity solutions are of interest, as the design is focused on 
the dynamic aperture optimization. In this case, it is thus preferable to chose a cell that can achieve an 
absolute minimum emittance much lower than the requirement of the design and detune it to large detuning factors (low phase advances), 
in order to minimize the chromaticity. A scan on the drift spaces lengths can then be performed in order to find the optimal configuration of the cell.
Here, the example of a detuned cell is considered for a dipole bending angle of $\theta=2\pi/38$, 
corresponding to a minimum emittance of 6.5~$\mu$m-rad and for a detuning factor of $\epsilon_r=10$. The emittance that 
this cell achieves, even if detuned by a factor of 10, is still within the requirements of the PDR design. 
Unlike the case of the absolute minimum emittance described in Sec.~\ref{sec:DriftParam}, in this case all choices of ($s_1$, $s_2$, $s_3$) 
triplets assure optical stability. Fig.~\ref{fig:driftscandetuned} shows 
only low chromaticity solutions for which $|\xi_{x,y}| < 2$. The 
parametrization of the horizontal chromaticity (left) and the horizontal dispersion at the middle of the dipole (right) 
with the ($s_1$, $s_2$, $s_3$) triplets is presented in the top part of the figure, while the projection to the ($s_1$, $s_2$) 
plane, color-coded with the horizontal chromaticity, in the bottom. There is a clear correlation between the  
$D_{x\mathrm{c}}$ and the horizontal chromaticity $\xi_x$. High chromaticity values correspond to negative or small dispersion 
at the middle of the dipole, as low dispersion values require strong focusing by the quadrupoles. 
Higher dispersion values at the middle of the dipole, correspond to smaller chromaticity values of the cell. 
Finally, the optimal configuration of the drift spaces lengths, for the example under study, in order to provide 
low chromaticity in a relatively short cell, was chosen to be ($s_1$, $s_2$, $s_3$)=(0.9, 0.6, 0.5)~m.

\subsection{Non linear optimization}
\label{sec:PDR-NonLinearOpt}

The main limitation of the DA in the low emittance lattices comes from the non-linear effects induced 
by the strong sextupole strengths, which are introduced for the chromaticity correction. 
From the non-linear dynamics theory~\cite{ref:nonlinearsls}, a resonance of order $n$ defined by $n_x Q_x+n_y Q_y=p$, with $|n_x|+|n_y|=n$ 
the order of the resonance and $p$ any integer, is associated with a driving term.  
Based on~\cite{ref:ResFreeLat}, the driving term of a resonance associated with the ensemble of $N_c$ cells vanishes, if the resonance amplification 
factor is zero:
\begin{equation}
 \label{eq:resonancecancellation}
  \left| \sum_{\mathrm{p=0}}^{\mathrm{N_c-1}} e^{\mathrm{ip(n_x\mu_{x,c}+n_y\mu_{y,c})}} \right|=
  \sqrt{\frac{1-\cos[N_c(n_x\mu_{x\mathrm{,c}}+n_y\mu_{y\mathrm{,c}})]}{1-\cos (n_x \mu_{x\mathrm{,c}}+n_y \mu_{y\mathrm{,c}})}}=0.
\end{equation}
This is achieved if: $N_c(n_x\mu_{x\mathrm{,c}}+n_y\mu_{y\mathrm{,c}})=2k\pi$, provided the denominator of 
Eq.~\eqref{eq:resonancecancellation} is non zero, i.e.: $n_x\mu_{x\mathrm{,c}}+n_y\mu_{y\mathrm{,c}}\neq 2k'\pi$, with $k$ and $k'$ 
any integers. 
From this, a part of a circular accelerator will not contribute to the excitation of any non-linear resonances, except
of those defined by $\eta_x \mu_x +\eta_y \mu_y = 2 k_3 \pi$, if the phase
advances per cell satisfy the conditions: $N_c \mu_x = 2k_1\pi$ and $N_c \mu_y = 2k_2\pi$, where $k_1$ , $k_2$ and $k_3$ are 
any integers. Prime numbers for $N_c$, which in our case is the number
of TME cells per arc, are interesting, as there are less resonances satisfying both diophantine conditions simultaneously.

The nonlinear optimization of the CLIC PDR lattice was based on the resonance free lattice concept, described above.
From Eq.~\eqref{eq:emitgen} and using a dipole field of $B_d$=1.2~T, at least 19 dipoles are needed 
in order to achieve the required output emittance. From this, convenient numbers of $N_c$ (number of 
dipoles per arc) are 11, 13 and 17, which means 26, 30 and 38 dipoles in the ring respectively, 
including the dispersion suppressors' dipole. 
Following the results from the analytical parametrization of the TME cells, small horizontal and vertical phase advances 
and large detuning factors are favorable, for low cell chromaticity. 
The largest number of cells is better for increasing
the detuning factor between the required and the minimum
emittance and the cancellation of a larger number of resonance driving terms. Finally, the option of $N_c=38$ 
was chosen. 

\begin{figure}[htp]
   \centering
    \includegraphics*[width=0.32\textwidth]{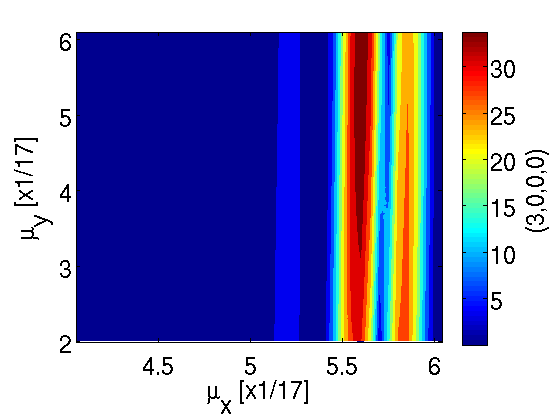}
    \includegraphics*[width=0.32\textwidth]{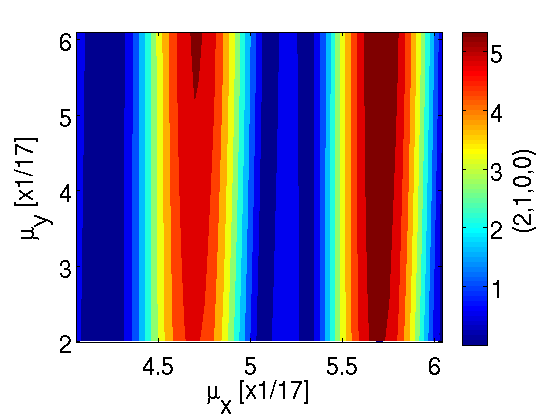}
    \includegraphics*[width=0.32\textwidth]{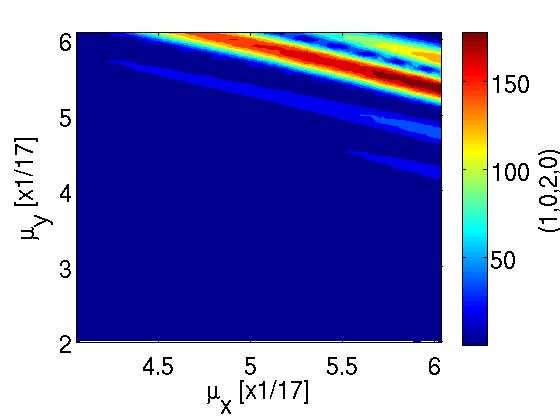}
    \includegraphics*[width=0.32\textwidth]{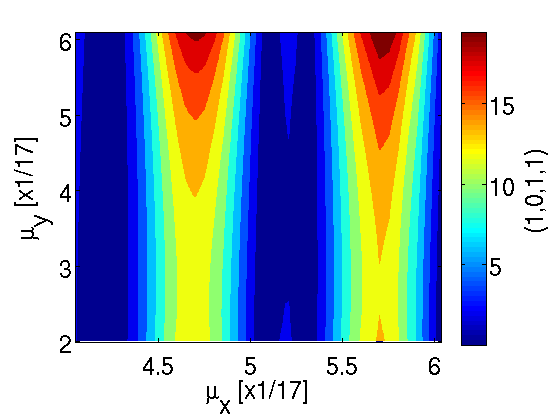}
    \includegraphics*[width=0.32\textwidth]{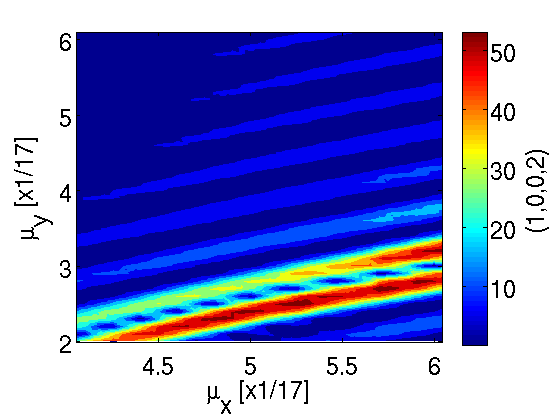}
    \caption{\label{fig:phaseadvscanHA3rd} Horizontal and vertical phase advances of the PDR TME cell, parametrized with the third order Hamiltonian amplitudes.}
   \vspace{20px}
\centering
    \includegraphics*[width=0.32\textwidth]{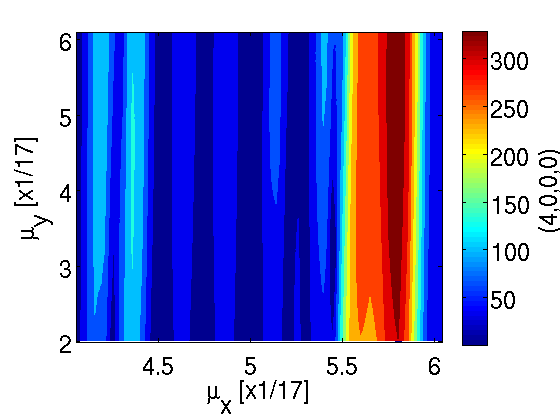}
    \includegraphics*[width=0.32\textwidth]{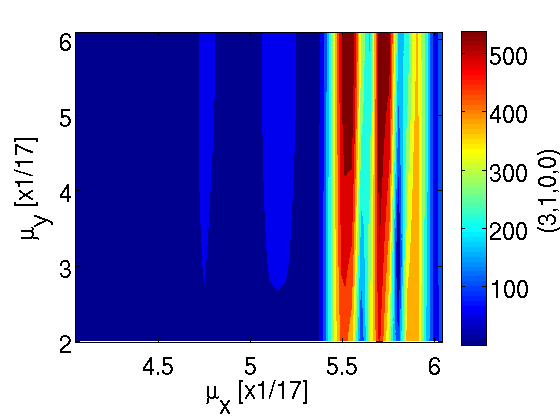}
    \includegraphics*[width=0.32\textwidth]{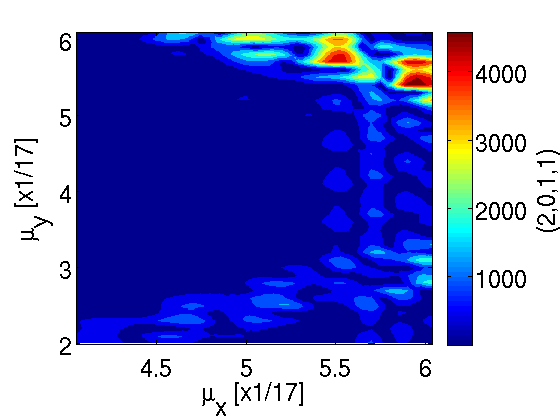}
    \includegraphics*[width=0.32\textwidth]{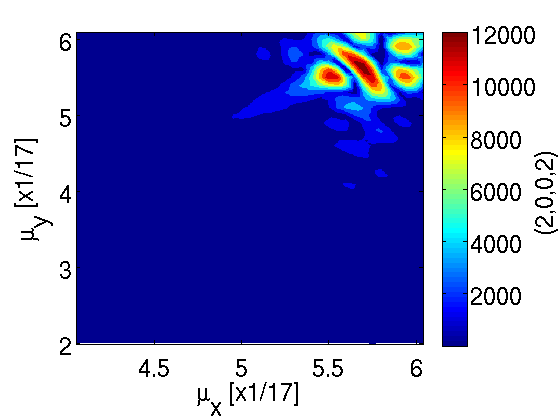}
    \includegraphics*[width=0.32\textwidth]{PDRpaper-FIG33.png}
    \includegraphics*[width=0.32\textwidth]{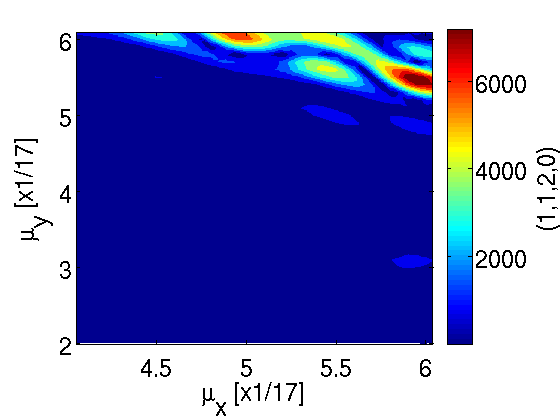}
    \includegraphics*[width=0.32\textwidth]{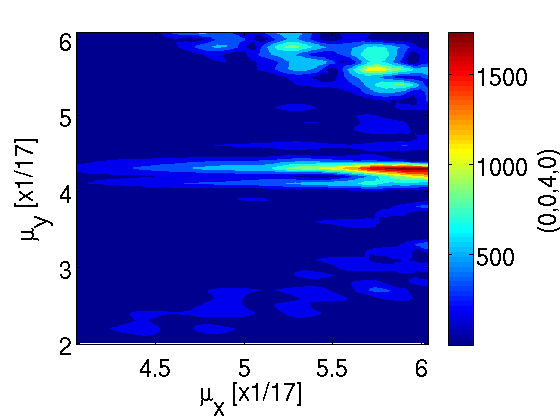}
    \includegraphics*[width=0.32\textwidth]{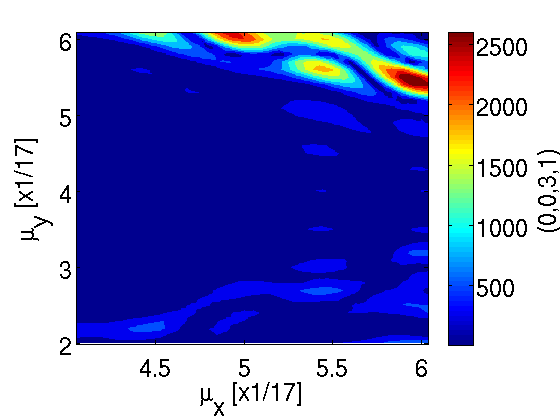}
     \caption{\label{fig:phaseadvscanHA4rth} Horizontal and vertical phase advances of the PDR TME cell parameterized with the fourth order Hamiltonian amplitudes.}   
\end{figure}

For the calculation of the resonance driving terms, the PTC-normal module of the MADX code is used, taking into account 
dipole and quadrupole fringe fields. The calculations are performed for different phase advances of the TME cell, 
while the resonance driving terms are calculated, for all the lattice. 
Fig.~\ref{fig:phaseadvscanHA3rd} shows the dependence of the third order resonance driving terms, for which 
$(j-k)\mu_x+(l-m)\mu_y=n$ and $|j-k|+|l-m|=3$, on the horizontal and vertical phase advances of the TME cell. 
Blue regions correspond to small resonance excitation, while red regions indicate maximum excitation. Comparing the 5 Hamiltonian resonant coefficients, 
the (2,1,0,0) mode is almost suppressed (a factor of 4-6 smaller than the other modes) and the (1,0,1,1) mode is weakly excited. 
The non-linear coupling term (1,0,2,0) is excited at high horizontal or high vertical phase advances. The horizontal mode 
(3,0,0,0) is also excited at high horizontal phase advances. In all cases, 
minimum excitation is observed, for integer multiples of 1/17.
  
Due to the fact that strong sextupoles are introduced in the PDR lattice, for the chromaticity correction, 
higher order resonances should also be considered.
Fig.~\ref{fig:phaseadvscanHA4rth} shows the dependence of the fourth order 
resonance driving terms, for which $|j-k|+|l-m|=4$, on the horizontal and vertical phase advances of the TME cell. 
Maximum excitation is observed for the non-linear coupling terms (2,0,0,2), (2,0,2,0), (1,1,2,0) and (2,0,1,1), especially at 
the high horizontal or high vertical phase advance limit of the scan. The horizontal modes (4,0,0,0) and (3,1,0,0) are weakly excited with respect 
to the other modes. The vertical modes (0,0,4,0) and (0,0,3,1) are also excited, in the high horizontal phase advance limit for the first 
and in the high vertical phase advance limit for the second. 
All resonance driving terms are suppressed, 
for phase advances that are integer multiples of 1/17, as expected.

Here, the resonance driving terms are presented and discussed only to demonstrate the proof of principle of the resonance 
free lattice concept. 
In a further non-linear optimization of the lattice, especially when high-order magnet errors are included, additional families 
of sextupoles, in non-dispersive areas, can be used for the minimization of the resonance driving terms which limit the dynamic aperture.

 \begin{figure}
    \centering
    \includegraphics*[width=0.45\textwidth]{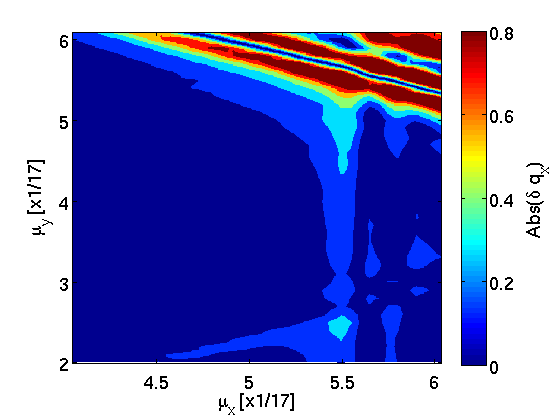}
    \includegraphics*[width=0.45\textwidth]{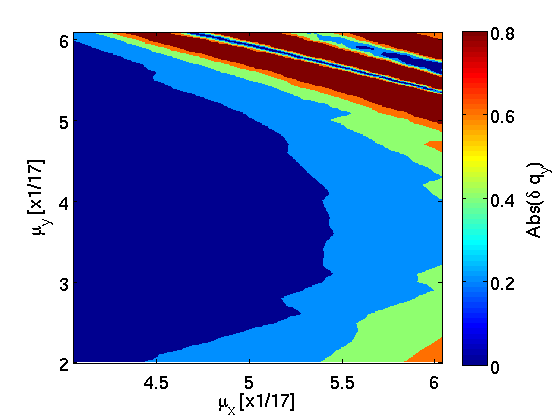}
    \includegraphics*[width=0.45\textwidth]{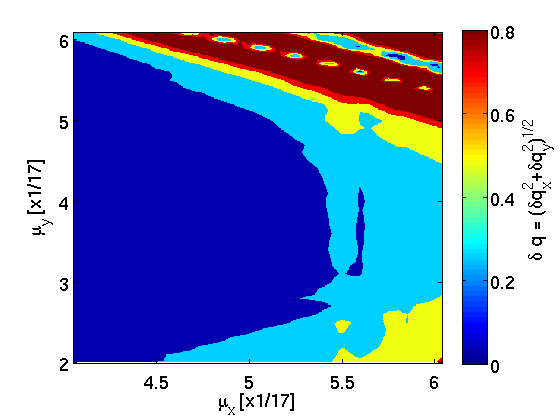}
    \includegraphics*[width=0.45\textwidth]{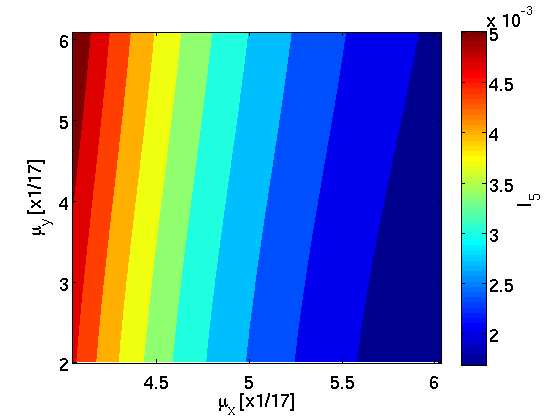}
     \caption{\label{fig:phaseadvscan} Top: Parametrization of the horizontal (top, left) and vertical (top, right) amplitude dependent
     tune shift, with the horizontal and vertical phase advances of the TME cell. 
     Bottom: Parametrization of the square root of the quadratic sum of the horizontal and vertical amplitude dependent
     tune shifts (left) and of the fifth radiation integral (right), with the horizontal and vertical phase advances of the TME cell.}    
\vspace{10px}
    \centering
    \includegraphics*[width=0.32\textwidth]{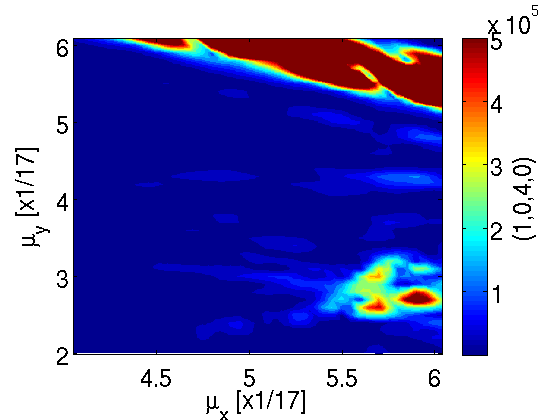}
    \includegraphics*[width=0.32\textwidth]{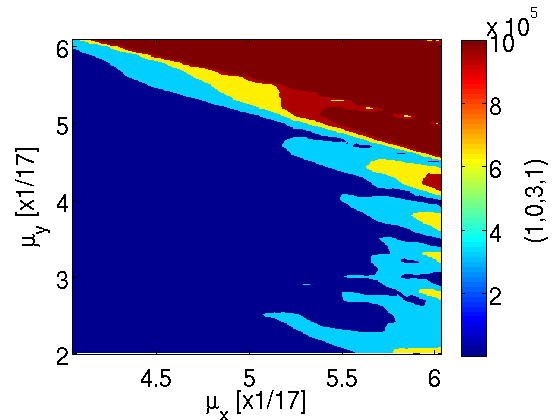}
    \includegraphics*[width=0.32\textwidth]{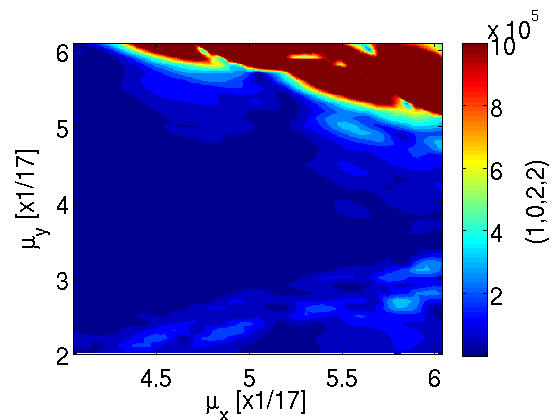}
    \includegraphics*[width=0.32\textwidth]{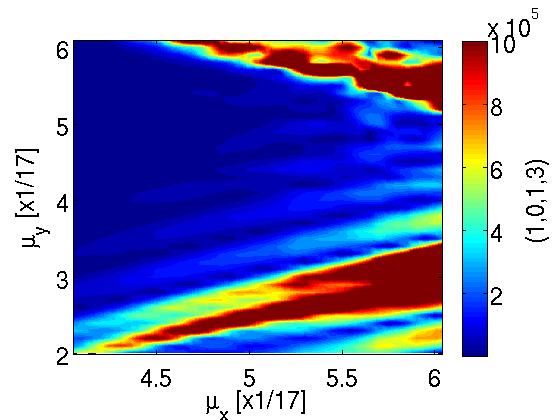}
    \includegraphics*[width=0.32\textwidth]{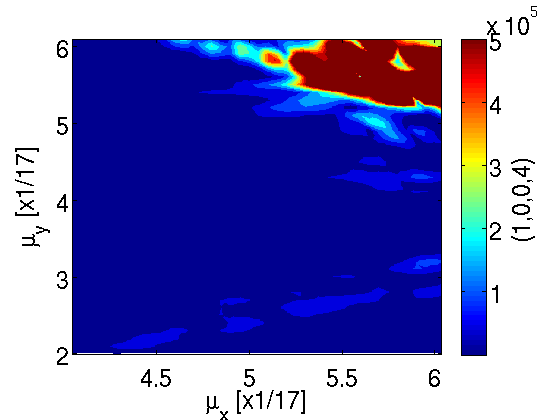}
     \caption{\label{fig:phaseadvscanHA5th} The fifth order resonance driving terms for which $|j-k|=1$ and $|l-m|=4$.}    
 \end{figure}
Another quantity that has to be taken into account, is the amplitude dependent tune shift 
$\delta q_{x,y}/\delta J_{x,y}$. From first order 
perturbation theory, the leading order tune shift can be represented by~\cite{ref:ForestFringeFields}:
\begin{equation}
\label{eq:dq}
\left( \begin{array}{c}
        \delta q_x \\
	\delta q_y
\end{array} \right) = 
\left( \begin{array}{ccc}
        \alpha_{\mathrm{hh}} & \alpha_{\mathrm{hv}} \\
	\alpha_{\mathrm{vh}} & \alpha_{\mathrm{vv}}  
\end{array} \right)
\left( \begin{array}{c}
         2 J_x \\
	 2 J_y
\end{array} \right),
 \end{equation}
where, $\alpha_{ij}$ are called the normalized anharmonicities and they describe the variation of the tune at different amplitudes 
(or action).

Fig.~\ref{fig:phaseadvscan} shows the dependence of the horizontal (top, left) and vertical (top, right) 
detuning with amplitude, $\delta q_x$ and $\delta q_y$ respectively, on the horizontal 
and vertical phase advances. 
The bottom plots show the parametrization of the factor 
$\delta q = \sqrt{\delta q_x^2+\delta q_y^2}$ (left) and the fifth radiation integral $I_5$ (right), 
which is an equivalent
to the horizontal emittance, with the horizontal and vertical phase advances.
The amplitude dependent tune shift gets larger for large phase advances, while the emittance follows the opposite behavior. 
For this reason a compromise solution is chosen, where the horizontal emittance is reached for a small (but not minimum) 
detuning with amplitude. The optimal solution was chosen to be $\mu_x$= 5/17 and $\mu_y$=3/17. With this choice, 
a compromise is achieved, for exciting the 
smallest number of resonances and achieve a rather small amplitude detuning and chromaticity, staying within 
the output emittance requirements of the design.

However, numerology shows that for this choice of phase advances, the non-linear fifth order coupling resonance driving terms 
are excited, for $|j-k|=1$ and $|l-m|=4$. In this case, $\mu_x+4\mu_y=5/17+4\times 5/17=1$. 
The five modes are presented in Fig.~\ref{fig:phaseadvscanHA5th}, with the (1,0,1,3) mode being the dominant for 
$\mu_x$= 5/17 and $\mu_y$=3/17. The other terms get excited for higher vertical phase advances.

For the chromaticity correction, four families of sextupoles are used. A set of sextupoles are located before the focusing 
quadrupoles of the TME cells and a set of sextupoles after the defocusing ones. The same set-up is followed for the 
two other families of sextupoles, which are placed in the half TME cells of the dispersion suppressors. As those sextupoles 
are not placed in dispersive areas, they do not contribute to the chromaticity correction, but they can be used for further 
non-linear optimization of the lattice. 

\begin{figure}[ht]
\centering
\includegraphics[width=0.5\linewidth,angle=270]{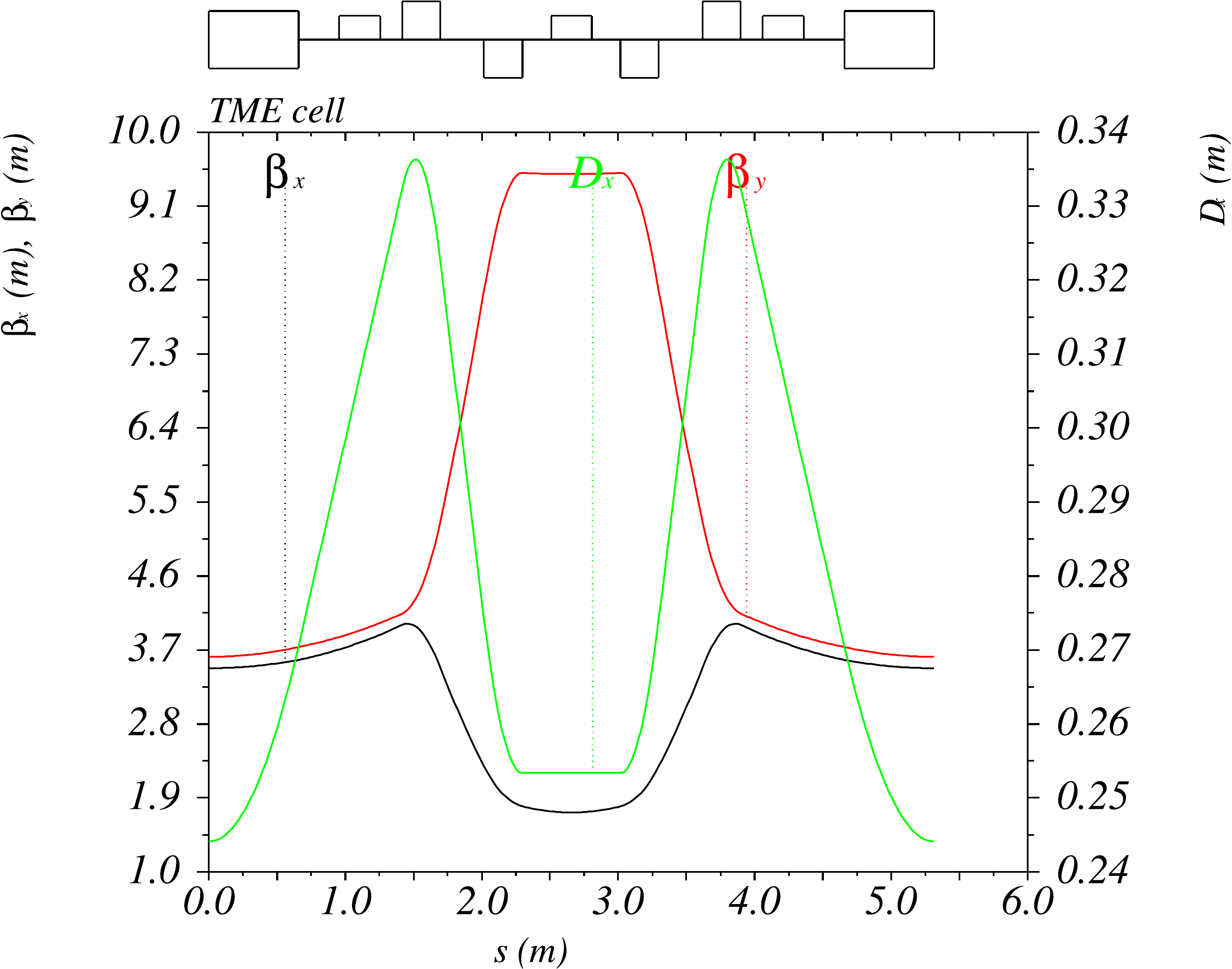}
\caption{The optical functions of the TME arc cell of the PDR.}
\label{fig:PDR-TME}
\end{figure}
The optical functions of the TME cell are shown in Fig.~\ref{fig:PDR-TME}, where the horizontal (black) and vertical (red) 
beta functions and the horizontal dispersion (green) along the cell are depicted. 

 \begin{figure}[ht]
 \centering
 \includegraphics[width=.5\linewidth]{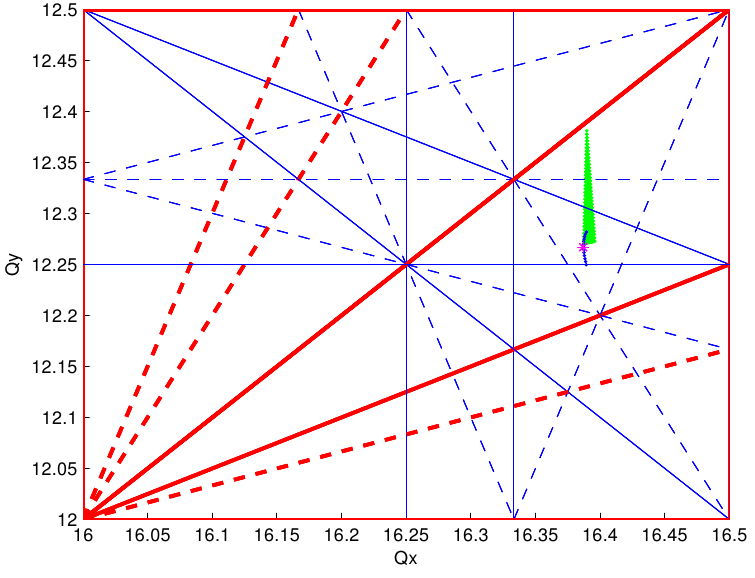}
 \caption{The working point in tune space for $\delta p/p$ from -1.2\% to 1.2\% (blue)
 	  and the first order tune shift with amplitude up to 6~$\sigma$ (green). The on-momentum working point is 
 	  (16.39, 12.27).}
 \label{fig:PDR-tune}
 \end{figure}
The change in the particles betatron frequencies, due to the non-linearities of the accelerator, can lead to 
the crossing of resonance lines in the tune diagram. This results 
in beam emittance blow up or in beam loss, thus, a careful choice of the betatron tunes of the linear lattice 
is very important for the beam quality and the beam life time.
In the CLIC PDR lattice, the betatron tunes are controlled by the quadrupoles of the long straight section 
FODO cells.  Fig.~\ref{fig:PDR-tune} shows the working point in tune space for momentum 
deviations $\delta p/p_0$ from -1.2$\%$ to 1.2$\%$ (blue) and the first order tune shift with amplitude 
(green) up to 6~$\sigma_{x,y}$. The on-momentum working point of the linear lattice is 
($Q_x$, $Q_y$)=(16.39,~12.26).

\subsection{Dynamic aperture}
\label{sec:PDR-DA}

The Dynamic aperture (DA) is defined as the maximum phase-space amplitude within which particles do not get lost as a consequence 
of single-particle effects~\cite{ref:Chao}. 
The DA has to be at least equal or larger than the minimum beam transverse acceptance, $R_{\mathrm{min}}$. 
The beam coming from the positron source is not expected to be Gaussian, and the distribution in the storage ring is not 
modified, until the beam is damped close to equilibrium. For this reason, the minimum transverse acceptance is defined 
in terms of a maximum emittance $\epsilon_{\mathrm{max}}$ of the particles with the maximum betatron action in the 
beam, and of a maximum relative momentum deviation $(\delta p/p_0)_{\mathrm{max}}$~\cite{ref:WolskiNLCPDR}: 
\begin{equation}
 R_{\mathrm{min}}=\sqrt{2 \beta \epsilon_{\mathrm{max}}}+D (\delta p/p_0)_{\mathrm{max}}.
\label{eq:Racc}
\end{equation}

\begin{figure}
\centering\includegraphics[width=.49\linewidth]{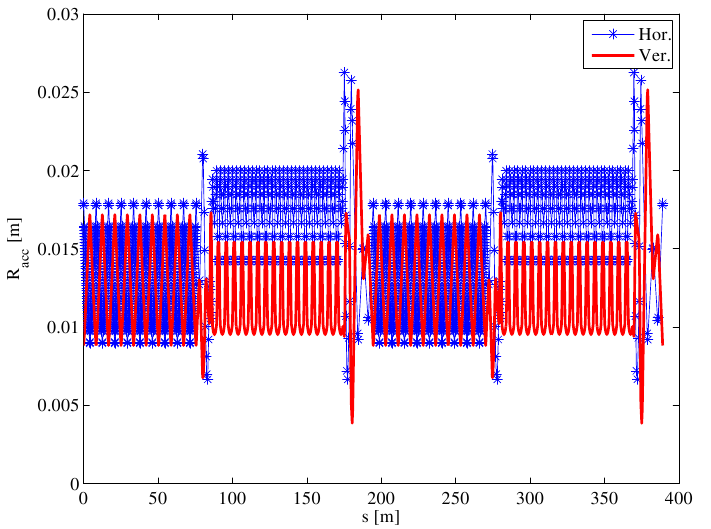}
\centering\includegraphics[width=.49\linewidth]{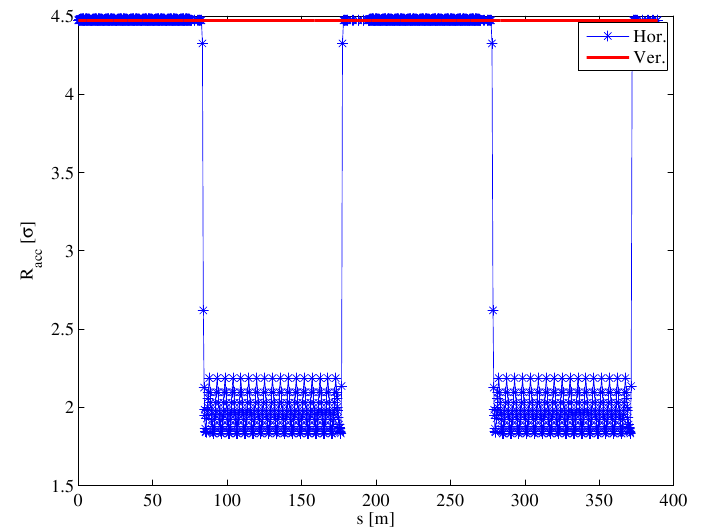}
\caption{The required acceptance around the PDR in order to fit the positron beam in units meters (left) and
in units of beam sizes (right).}
\label{fig:PDR-Accept}
\vspace{15px}
\end{figure}
The incoming beam to the CLIC PDR is a round beam with same horizontal 
and vertical rms emittances of $\epsilon^{\mathrm{rms}}_{x,y}$=7~mm-rad where, 99.9~\% of the particles are inside a maximum emittance of 
$\epsilon_{\mathrm{max}}$=10$\epsilon^{\mathrm{rms}}_{x,y}$ and with maximum $(\delta p/p_0)_{\mathrm{max}}=3\%$. Applying this to 
Eq.~\ref{eq:Racc}, the minimum acceptance can be calculated around the ring and is shown in Fig.~\ref{fig:PDR-Accept}, 
in units of [m] (left) and in units of beam sizes [$\sigma$] (right).
A minimum DA of 4.5$\sigma_{x,y}$ is required, in both horizontal (blue) and vertical (red) planes, 
in order to fit the large non-Gaussian beam coming from the positron source.

\begin{figure}[ht]
\centering
\includegraphics[width=.6\linewidth]{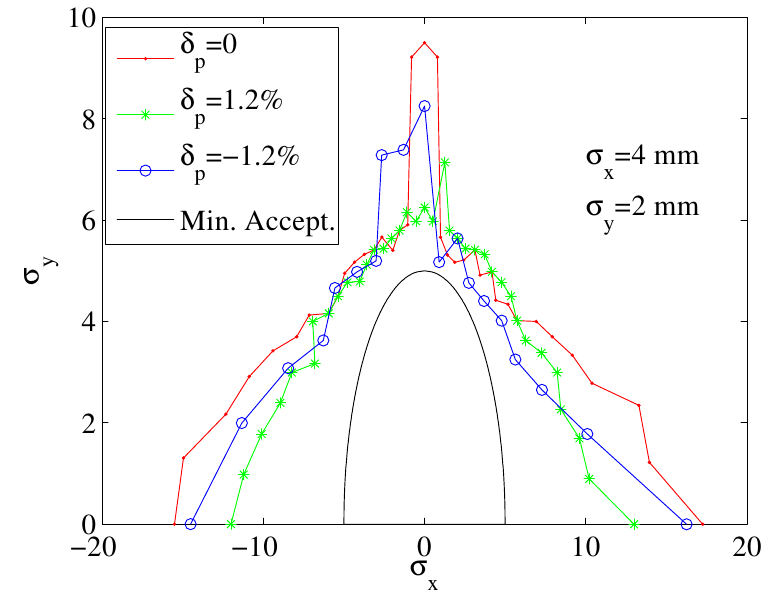}
\caption{The on and off momentum Dynamic Aperture of the PDR for $\delta_p=0$ (red), 1.2~\% (green) and 
	   -1.2~\% (blue).}
\label{fig:PDR-DA}
\end{figure}
The DA of the ring was computed with numerical particle tracking, over 
1000 turns, with the PTC module of MADX~\cite{bib:MADX}. 
Fig.~\ref{fig:PDR-DA} shows the initial positions of particles that 
survived over 1000 turns, normalized to the horizontal and vertical 
beam sizes, at the point of calculation ($\sigma_x$=4~mm, $\sigma_y$=2~mm). The results for $\delta p/p_0=0\%$ are shown in red, for $\delta p/p_0=1.2\%$ in 
green and for 
$\delta p/p_0=-1.2\%$ in blue. The minimum acceptance is shown in black. 
For these calculations the magnet fringe fields are taken into account, while any magnet error effects are 
neglected. An adequate but tight dynamic aperture 
is demonstrated, following an optimization procedure based on the resonance free lattice concept, however, 
more optimization steps is required when magnet errors and the effect of wigglers are included.

\subsection{Frequency maps}
\label{sec:PDR-FM}

\begin{figure}[pht]
\centering\includegraphics[width=.45\linewidth,height=.32\linewidth]{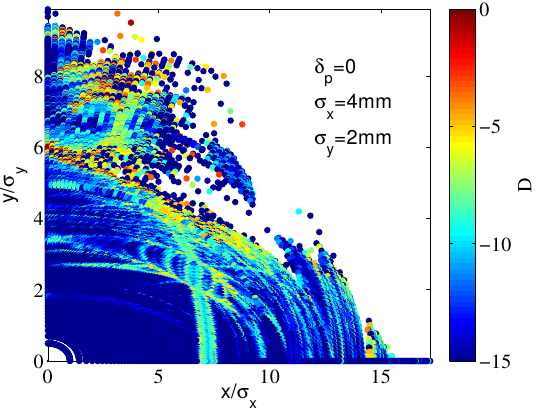}
\centering\includegraphics[width=.45\linewidth,height=.32\linewidth]{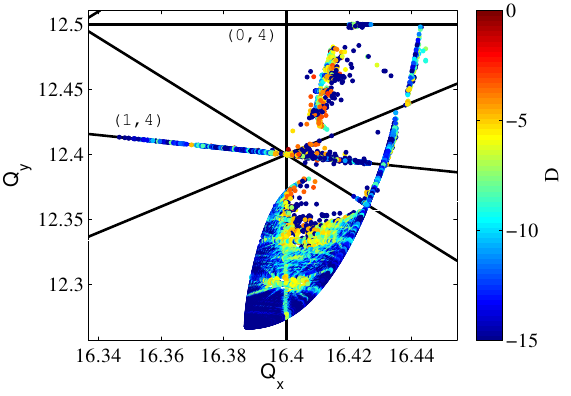} \\
\centering\includegraphics[width=.45\linewidth,height=.32\linewidth]{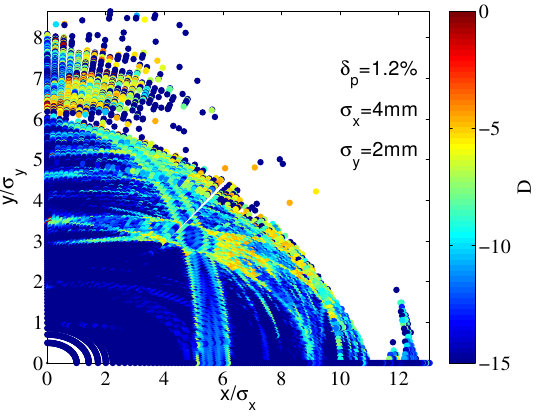}
\centering\includegraphics[width=.45\linewidth,height=.32\linewidth]{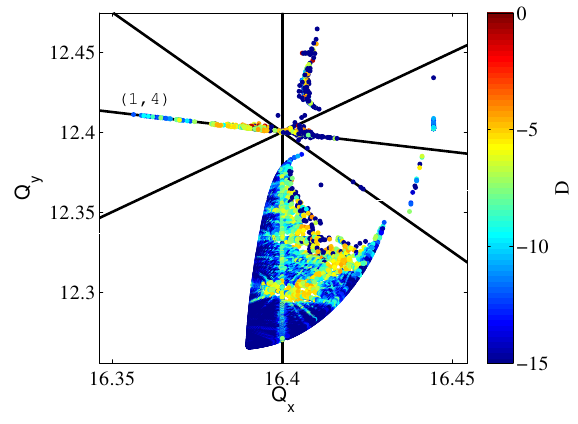} \\
\centering\includegraphics[width=.45\linewidth,height=.32\linewidth]{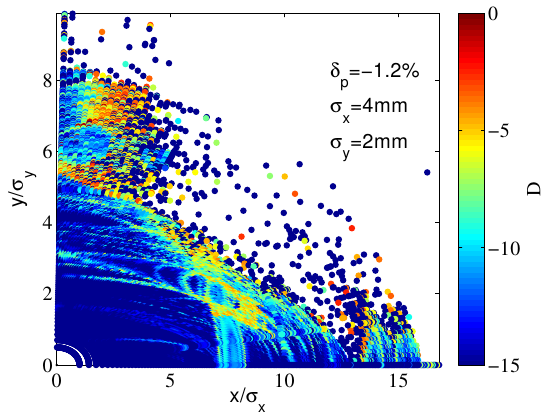} 
\centering\includegraphics[width=.45\linewidth,height=.32\linewidth]{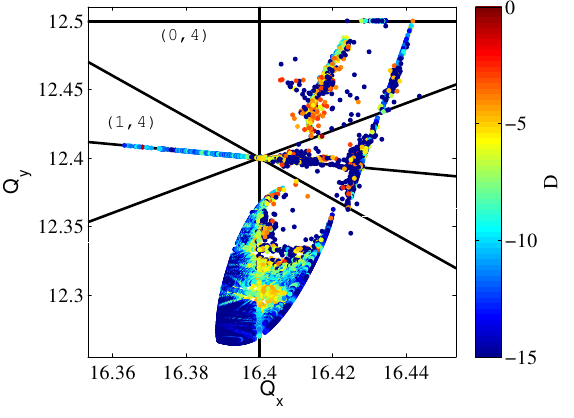}
\vspace{-15px}
\caption{Diffusion maps (left) and frequency maps (right) for $\delta p/p$=0 (top), 1.2\% (middle) and -1.2\% (bottom).}
\vspace{15px}
\label{fig:FM-PDR}
\end{figure}

The frequency map analysis (FMA) examines the dynamics in frequency space rather than
configuration space. Regular or quasi-regular periodic motion is a single point in 
frequency space
characterized by a pair of fixed tune values.
Irregular trajectories exhibit diffusion in frequency space, with the tunes changing
in time.
The mapping of configuration space ($x$ \& $y$) to frequency space ($Q_x$ \& $Q_y$) will be
regular for regular motion and irregular for chaotic motion.
Numerical integration of the equations of motion, for a set of initial
conditions ($x$, $y$, $x'$, $y'$) and computation of the frequencies as a function of time (or turn
number), constructs the map from the space of initial
conditions to frequency or tune space, over a finite time span T~\cite{ref:NAFF,ref:NAFF2,ref:NAFF3}. 
An indication of how much the frequency is changing with time, is measured through 
the diffusion coefficient, defined by:
\begin{equation}
 {\cal D}=\log\sqrt{(Q_{x\mathrm{1}}-Q_{x\mathrm{2}})^2+(Q_{y\mathrm{1}}-Q_{y\mathrm{2}})^2}
\label{eq:FMdiff}
\end{equation}
where the index 1 refers to a certain number of turns, while, the index 2  to a consecutive same amount of turns.
Large negative values of ${\cal D}$ denote long term stability while values of ${\cal D}$ close to zero denote 
chaotic motion~\cite{ref:NAFF}.

Tracking of particles with different initial conditions for 1024 turns, was performed with MADX-PTC~\cite{ref:madx-ptc}.
The ideal lattice including sextupoles and fringe fields is used, while no magnet errors are taken into account. 
The frequency map analysis was performed with the Numerical Analysis of Fundamental Frequencies (NAFF) algorithm~\cite{ref:NAFF}.

Fig.~\ref{fig:FM-PDR} (left) shows the initial positions of particles survived over 1024 turns, color-coded with the 
diffusion coefficient of Eq.~\eqref{eq:FMdiff}, for on-momentum particles with $\delta p/p_0$=0\% (top) and for 
off-momentum particles with $\delta p/p_0$=1.2\% (middle) and $\delta p/p_0$=-1.2\% (bottom). The particle positions in the horizontal and vertical axis 
are expressed in units of horizontal and vertical beam sizes at the point of calculation, where $\sigma_x$=4~mm 
and $\sigma_x$=2~mm. The frequencies of the particles are presented in the right plots, the frequency maps.
The color indicates the regularity of the orbits. Blue regions indicate very regular motion, while dark-red region 
indicate chaotic motion. The absence of dots means that the particles were lost. Resonance lines in the frequency maps
are shown as distorted areas, while the colors allow to relate the resonant features observed, to regions of the physical 
space~\cite{ref:NAFF}. 
From the frequency maps it is observed that the tune is crossing the (1,4) resonance, which is not eliminated 
by the TME phase advance choice ($\mu_x=5/17,~\mu_y=3/17$) as shown in Fig.~\ref{fig:phaseadvscanHA5th}. 
This seems to be the main limitation of the DA. 

The shape of the frequency maps, especially at high amplitudes, 
does not have the triangular shape expected by the linear dependence of the tune shift to the action, and 
they appear to be folded. This occurs when terms of higher order in the
Hamiltonian become dominant over the quadratic terms as the amplitude increases~\cite{ref:NAFF}. 
This behavior occurs due to the suppression of the lower order resonances, following the resonance free lattice 
concept, which gives rise to higher order terms. Even though folded maps may lead to potentially very
unstable designs, in our case this is not taken into account for the moment, as the folding of the map appears at 
high amplitudes, beyond the DA aperture limit. 

\section{\label{sec:Conclusion}Conclusion}

An analytical parametrization for the TME cell has been derived and presented in this paper, based on linear optics 
arguments and the thin lens approximation. In that way all cell properties, optical and geometrical, are globally 
determined and the optimization procedure following any design requirements can be performed in a systematic way. 
Stability criteria in both horizontal and vertical planes and magnet technology constraints are also applied. A 
comparison of the analytical solution with the results from the simulation code MADX gave very good agreement, 
even for the thick-elements optics. This method provides a very useful tool for defining optimal regions of 
operation for the best performance of the cell, according to the requirements of the design. 
The analytical approach and the resonance free lattice concept were finally used for the linear and non-linear 
optimization of the CLIC Pre-damping rings, providing an adequate dynamic aperture for a large incoming beam.

\newpage 
\bibliography{TMEpaper}

\end{document}